\documentclass[aip,graphicx]{revtex4}

\pdfoutput=1

\usepackage{color}
\usepackage{graphics}
\usepackage[pdftex]{graphicx}
\usepackage{amsmath}
\usepackage{amssymb}
\usepackage{amsfonts}
\usepackage{float}
\usepackage{natbib}
\usepackage{epstopdf}
\usepackage{verbatim}
\usepackage{algorithm}
\usepackage{algorithmic}
\usepackage{todonotes}
\usepackage{lineno}
\usepackage{bbold}
\usepackage{array}
\usepackage{mathrsfs}
\usepackage{amsbsy}
\usepackage{mathtools}
\usepackage{chngcntr}
\usepackage{cleveref}

\newcolumntype{?}{!{\vrule width 1pt}}
\newcolumntype{C}{>{\centering\arraybackslash}p{3.7cm}}
\newcolumntype{L}{>{\raggedright\arraybackslash}p{3.9cm}}
\newcolumntype{e}{>{\raggedright\arraybackslash}p{2.5cm}}
\newcolumntype{k}{>{\raggedright\arraybackslash}p{2cm}}
\newcolumntype{l}{>{\raggedright\arraybackslash}p{3.2cm}}
\newcolumntype{R}{>{\raggedleft\arraybackslash}p{3.7cm}}

\newcommand{\Rey}{Re}
\newcommand{\thalf}{\tfrac{1}{2}}
\newcommand{\squart}{\tfrac{1}{4}}
\newcommand{\bnabla}{\mbox{\boldmath$\nabla$}}
\newcommand{\bcdot}{\mbox{\boldmath$\cdot$}}
\newcommand{\mathsfbi}[1]{\ensuremath{\mathbf{#1}}}

\begin{document}

\title{Computational mechanics of soft filaments}

\title{Streaming enhanced flow-mediated transport}

\author{Tejaswin Parthasarathy}
\affiliation{Mechanical Sciences and Engineering, University of Illinois at Urbana-Champaign, Urbana, IL 61801, USA}

\author{Fan Kiat Chan}
\affiliation{Mechanical Sciences and Engineering, University of Illinois at Urbana-Champaign, Urbana, IL 61801, USA}

\author{Mattia Gazzola}
\email{mgazzola@illinois.edu}
\affiliation{Mechanical Sciences and Engineering and National Center for Supercomputing Applications, University of Illinois at Urbana-Champaign, Urbana, IL 61801, USA}


\begin{abstract}
We investigate the capability of an active body (master) to manipulate a passive object (slave) purely via contactless flow-mediated mechanisms, motivated by potential applications in microfluidic devices and medicine (drug delivery purposes). We extend prior works on active--passive cylinder pairs by superimposing periodic oscillations to the master's linear motion. In a viscous fluid, such oscillations produce an additional viscous streaming field, which is leveraged for enhancing slave transport. We see that superimposing oscillations robustly improves transport across a range of Reynolds numbers. Comparison with results without oscillations highlights the flow mechanisms at work, which we capitalize on to design (master) geometries for augmented transport. These principles are found to extend to three-dimensional active--passive shapes as well.
\end{abstract}


\maketitle

\vspace{-10pt}
\section{Introduction}
\vspace{-5pt}
This paper considers two- and three-dimensional flow-mediated transport systems operating in flow regimes characterized by finite, moderate Reynolds numbers (\(1 \le \Rey \le 100 \)). In particular, we explore strategies based on viscous streaming effects (enabled by gentle oscillations) to enhance the capability of an active leading object to transport, trap and manipulate passive trailing ones.

We are motivated by the accelerated pace of development of artificial and biohybrid \citep{Williams:2014, Park:2016, Ceylan:2017} mini-bots enabled by recent simulation \citep{Gazzola:2018, Pagan-Diaz:2018} and fabrication advances \citep{Ceylan:2017}. This new breed of mini-bots predominantly operates in fluids, and brings within reach a range of novel high-impact applications in medicine and manufacturing (drug delivery and particle transport, chemical mixing and \textit{in-situ} contactless manipulation, among many others). Fluid-mediated interactions can then be leveraged to enhance these capabilities or enable new ones \citep{Ceylan:2017}.

In general flow-mediated interactions play an important role in a number of physical and biological phenomena, from fish schooling \citep{Weihs:1973} and suspension of microorganisms \citep{Ishikawa:2006,Koch:2011} to cloud particle sedimentation \citep{Metzger:2007} and cluster formation \citep{Voth:2002}. Thus, hydrodynamically-coupled systems have been investigated for different flow regimes, including Stokes (\( \Rey \to 0 \),~\cite{Brady:1988,Lauga:2009}), Oseen (\( \Rey \sim \textit{O}(1) \),~\cite{Subramanian:2008}) and inviscid (\( \Rey \to \infty \),~\cite{Nair:2007,alben2009wake,Tchieu:2012}). Moreover, interest in fish schooling also prompted a number of studies (\(\Rey > 100\),~\cite{Liao:2003,Liao:2007,Ristroph:2008,Borazjani:2008,Bergmann:2011,Tchieu:2012,Boschitsch:2014,Gazzola:2014,Gazzola:2015a}) including attempts to bridge viscous and inviscid descriptions \citep{Eldredge:2010}.

There has been instead little effort in characterizing flow coupling mechanisms in the range \(1 \le \Rey \le 100 \), typical of the emergent technologies highlighted above \citep{Ceylan:2017}. Yet, in this regime, systems of moving objects exhibit rich dynamics characterized by sharp transitions~\citep{Gazzola:2012a,Lieu:2012,lin2018effects}, leading to drastically different behaviours (transport vs.~non transport, attraction vs.~repulsion) depending on the ratio between viscous and inertial effects.

Given the sharpness of these transitions, we hypothesize that controlled perturbations or second-order effects such as viscous streaming can be leveraged to shift the boundaries between qualitatively different system responses in a rational, regulated fashion. Viscous streaming arises when a fluid of viscosity \( \nu \) is driven periodically with frequency \(\omega \) by a vibrating boundary of characteristic length \( D \), and is responsible for the generation of steady flow structures that exhibit large spatial and slow temporal scales (relative to \( D \) and \(\omega \)). This phenomenon is well understood, both theoretically and experimentally, in the case of individual vibrating cylinders \citep{Holtsmark:1954, Stuart:1966, Riley:1967, Davidson:1972, Riley:2001, Lutz:2005} and spheres \citep{Lane:1955,Riley:1966,Chang:1994}, and has found application in microfluidic flow control, mixing, sorting and pumping \citep{Marmottant:2003,Marmottant:2004,Lutz:2006a,He:2011,Liu:2002,Ahmed:2009,Wang:2011,Thameem:2016,Ryu:2010,Tovar:2011}. Yet little is known in the case of complex geometries or configurations involving multiple objects.

In this work we characterize the impact of viscous streaming in the context of passive two- and three-dimensional particle transport by capitalizing on our previous work \citep{Gazzola:2012a}~(\cref{fig:framework}a), and drawing inspiration from previous studies \citep{Tchieu:2010a,Chong:2013}. We thus consider the simple yet representative setting of~\cref{fig:framework}, characterized by a larger cylindrical bot (master) propelling at a constant forward speed and oscillating transversely, and a smaller, passive, trailing cargo (slave). By means of this idealized experiment, we investigate the slave's response, dissect the mechanisms at play and challenge our insights to design master geometries that improve transport.

This work is organized as follows: setup is detailed in $\S$\cref{sec:setup}; numerical method and validation are illustrated in $\S$\cref{sec:numerics}; two-dimensional transport applications, analyses and design are presented in $\S$\cref{sec:results}; extension to three dimensions is discussed in $\S$\cref{sec:3d}; findings are summarized in $\S$\cref{sec:conclusion}.

\vspace{-10pt}
\section{Physical framework and streaming definitions}\label{sec:setup}
\begin{figure}[h!]
\begin{center}
\includegraphics[width=\linewidth]{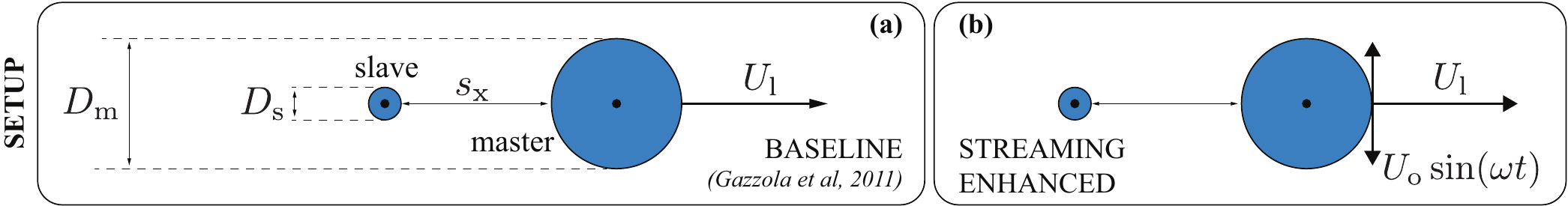}
\vspace{-20pt}
\caption{System setup: (\textit{a}) baseline~\citep{Gazzola:2012a} vs. (\textit{b}) current configuration.}
\label{fig:framework}
\end{center}
\vspace{-15pt}
\end{figure}

\vspace{-5pt}
We adopt the setup of~\cref{fig:framework}a with master and slave cylinders of diameter \(D_{\textrm{\scriptsize m }}\) (and radius \( r_{\textrm{\scriptsize m}} \)) and \( D_{\textrm{\scriptsize s}} = \squart D_{\textrm{\scriptsize m}}\), respectively. The slave is initially at rest, located at a separation distance \( 0.1D_{\textrm{\scriptsize m}} \) behind the master, which impulsively starts translating horizontally with a constant speed \( U_{\textrm{\scriptsize l}} \), spanning \( 1 \le \Rey \le 100 \). We refer to this system setup as the `baseline' throughout, chosen for its minimal complexity and consistency with prior works \citep{Tchieu:2010a,Gazzola:2012a}, thus aiding analysis and comparison. In~\cref{fig:framework}b we superpose to the linear motion of the master a transverse low-amplitude sinusoidal oscillation defined by \( y^{\textrm{\scriptsize m}}(t)= y^{\textrm{\scriptsize m}}(0) + \epsilon r_{\textrm{\scriptsize m}} \sin(\omega t)\) with characteristic velocity \(U_{\textrm{\scriptsize o}} = \epsilon \omega r_{\textrm{\scriptsize m}}\), where \( \omega \) is the angular frequency and \( \epsilon = A/r_m = 0.1 \) is the non-dimensional amplitude (\(A\) denotes amplitude). Oscillations elicit a viscous streaming response which can exert small but non-negligible forces, when compared against wake forces (Supplementary Material). We claim that this contribution, when properly directed, can significantly alter transport behaviour in a flow regime characterized by sharp transitions~\citep{Gazzola:2012a}, through constructive effects between wake and streaming components as hinted to in literature \citep{kubo1980secondary}.

We characterize the linear motion dynamics by \( \Rey =  U_{\textrm{\scriptsize l}} D_{\textrm{\scriptsize m}}/\nu \) and the oscillatory dynamics by $ R_{\textrm{\scriptsize o}} = U_{\textrm{\scriptsize o}} r_{\textrm{\scriptsize m}}/\nu$, where $\nu$ is the kinematic viscosity. We further define $ \zeta = R_{\textrm{\scriptsize o}}/\Rey  $ as the non-dimensional quantity that encodes the relative time scales of oscillatory and linear motion. Following \citep{Stuart:1966}, we characterize streaming dynamics through the streaming Reynolds number \( R_{\textrm{\scriptsize s}} = U_{\textrm{\scriptsize o}}^2/\nu\omega \), based on the oscillatory Stokes boundary layer length scale \( \sqrt{\nu t} = \sqrt{\nu/\omega}\) as \(t \sim \omega^{-1}\). While the magnitude of \(\Rey\) is an indicator of the flow regime due to linear motion, the magnitude of \(R_{\textrm{\scriptsize s}}\) similarly characterizes steady streaming structures. \Cref{fig:validation_panel}a,b (related to the validation of our numerical solver) illustrates these structures for an oscillating (with no linear translation) cylinder, which consists of distinct regions of clockwise~(blue) and counter-clockwise~(orange) rotating vortical fluid. \Cref{fig:validation_panel}a depicts a case with \(R_{\textrm{\scriptsize s}} \ll 1\), where viscous effects dominate and the steady streaming flow is Stokes-like, with a single large (infinite for practical purposes) boundary layer that helps match the no-slip condition on the body surface (indicating slow velocity decay~\citep{Riley:1967}). \Cref{fig:validation_panel}b is representative of \(R_{\textrm{\scriptsize s}}\) spanning \(\textit{O}(1)\)--\textit{O}\((10)\), where both inertial and viscous effects are comparable. This leads to an `outer' boundary layer (or DC boundary layer), which has a finite thickness depending only on the Womersley number \(\alpha = r_{\textrm{\scriptsize m}}\sqrt{\omega/\nu}\) (\cref{fig:validation_panel}c). Over this thickness, the fluid velocity adjusts to match the no-slip condition on the body surface. The DC boundary layer then `drives' the fluid in the bulk (from its outer surface) to set up long range streaming forces. We note that for fixed \(\epsilon\), as in this work, we can use \(R_{\textrm{\scriptsize o}}\) alone to characterize streaming flows (as \(R_{\textrm{\scriptsize o}} = R_{\textrm{\scriptsize s}}/\epsilon = \epsilon \alpha^2\ = \epsilon \omega r_{\textrm{\scriptsize m}}^2/\nu \)).

After introducing viscous streaming in this simple setting, we proceed to study the case of~\cref{fig:framework}b, through numerical simulations via remeshed vortex method~\citep{Gazzola:2011a}, briefly recapped in the following section.

\vspace{-10pt}
\section{Governing equations, numerical method and validation}\label{sec:numerics}
\vspace{-5pt}
We consider incompressible viscous flows in an infinite domain ($\Sigma$) in which two density-matched moving rigid bodies are immersed. We denote with $\Omega_i$ \& $\partial\Omega_i$ ($i=1,2$) the support and boundaries of the solids. Then the flow is described by the incompressible Navier--Stokes equations \cref{eq:simpleNS01}
\begin{equation}
\bnabla \bcdot \boldsymbol{u} = 0, \qquad
\frac{\partial \boldsymbol{u}}{\partial t} + \left( \boldsymbol{u} \bcdot \bnabla \right)\boldsymbol{u} = -\frac{1}{\rho}\nabla P + \nu \bnabla^2 \boldsymbol{u}~~~~\boldsymbol{x}\in\Sigma\setminus\Omega_i
\label{eq:simpleNS01}
\end{equation}
where $\rho$ is the fluid density. We have the no-slip boundary condition $ \boldsymbol{u} =  \boldsymbol{u}_i$ at the body--fluid interface $ \partial \Omega_i $, where $ \boldsymbol{u}_i$ is the \( i^{\textrm{\scriptsize th}}\) body velocity. The feedback from the fluid to the \( i^{\textrm{\scriptsize th}}\) body is described by Newton's equation of motions  $m_i \ddot{\boldsymbol{x}}_i = \boldsymbol{F}^H_i$ and $ \mathsfbi{I_i}\ddot{\boldsymbol{\theta_i}}=\boldsymbol{M}^H_i$, where $\boldsymbol{x}_i$, $\boldsymbol{\theta_i}$, $m_i$, $\mathsfbi{I_i}$, $\boldsymbol{F}^H_i$ and $\boldsymbol{M}^H_i$are, respectively, the position of the center of mass, angular orientation, mass, moment of inertia matrix, hydrodynamic force and moment. This system of equations is solved in the velocity--vorticity form by combining remeshed vortex methods with Brinkmann penalization and a projection approach \citep{Gazzola:2011a}. This method has been extensively validated across a range of biophysical problems, from bluff body flows to biological swimming \citep{Gazzola:2011a,Gazzola:2011,Gazzola:2012,Gazzola:2014}.

\begin{figure}[h!]
	\centerline{\includegraphics[width=\linewidth]{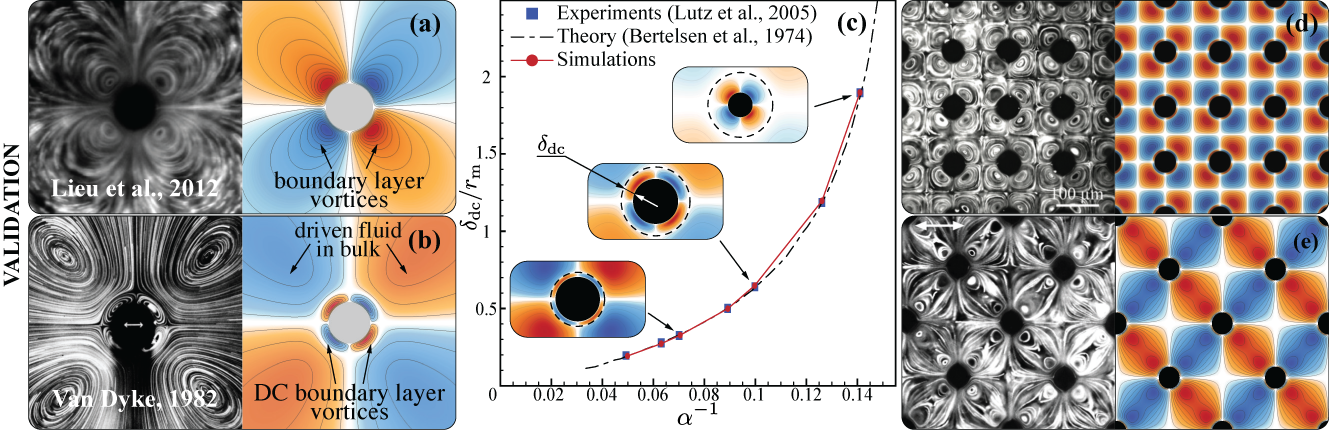}}
	\vspace{-10pt}
	\caption{Validation: Comparison of time-averaged streamline pattern in (\textit{a}) Stokes-like ($R_{\textrm{\scriptsize o}} = 0.8$) and (\textit{b}) Double Boundary Layer (DBL) ($R_{\textrm{\scriptsize o}} = 6.28$) regimes against the experiments of \citep{Lieu:2012} and \citep{vanDyke:1982}, respectively. (\textit{c}) Normalized DC boundary layer thickness $ \delta_{\textrm{\scriptsize dc}}/r_{\textrm{\scriptsize m}} $ vs. $ \alpha^{-1}$ against experiments \citep{Lutz:2005} and theory \citep{Bertelsen:1973} in the DBL regime. (\textit{d,e}) Multiple streaming cylinders comparison against the experiments of~\citep{House:2014}. Simulation details: domain $[0,1]^2~\textrm{m}^2$, uniform grid spacing $h=1/2048~\textrm{m}$, penalization factor $\lambda=10^4$, mollification length $\epsilon_{\textrm{\scriptsize moll}} = 2 \sqrt{2}h$, lagrangian CFL $=0.01$, with viscosity \(\nu\) and oscillation frequency \(\omega\) set according to prescribed linear (\(\Rey\)) and oscillatory (\(R_{\textrm{\scriptsize o}}\)) Reynolds numbers. The above values are used throughout the text, unless stated otherwise. We refer to \citep{Gazzola:2011a} for details on these parameters.}
	\label{fig:validation_panel}
	\vspace{-20pt}
\end{figure}

We now extend this validation to show that the algorithm can also accurately capture second-order streaming dynamics. In \cref{fig:validation_panel}a,b, we qualitatively compare the streaming structures in the Stokes-like~(\(R_{\textrm{\scriptsize s}} \ll 1\)) and Double Boundary Layer (DBL, with \(R_{\textrm{\scriptsize s}}\) spanning \(\textit{O}(1)\)--\(\textit{O}(10)\)) regimes against corresponding experiments. \Cref{fig:validation_panel}c shows the quantitative comparison against prior theory and experiments in the DBL regime, wherein we relate the DC boundary layer thickness (\( \delta_{\textrm{\scriptsize dc}} \), defined as the offset of the stagnation streamline from the cylinder surface, marked with dashed lines in the figure's insets) to the inverse Womerseley number \(\alpha^{-1} = \sqrt{\nu/\omega}/r_{\textrm{\scriptsize m}}\). We depict streaming structures from multiple oscillating cylinders in \cref{fig:validation_panel}d,e to highlight similarity with corresponding experiments. Overall, we observe qualitative and quantitative agreement with analytical and experimental studies involving individual and multiple cylinders, thus demonstrating the viability of our numerical approach to this class of problems.

\section{Transport in two dimensions}\label{sec:results}
\vspace{-10pt}
We now assess the impact of streaming on slave transport and quantify it by defining the non-dimensional surface-to-surface separation distance \( s_{\textrm{\scriptsize x}}(T)/D_{\textrm{\scriptsize m}} \) (\cref{fig:framework}) in the \( x \) direction.

\vspace{-15pt}
\subsection{Comparison with baseline}\label{ss:baseline}
\begin{figure}[h!]
	\centerline{\includegraphics[width=\linewidth]{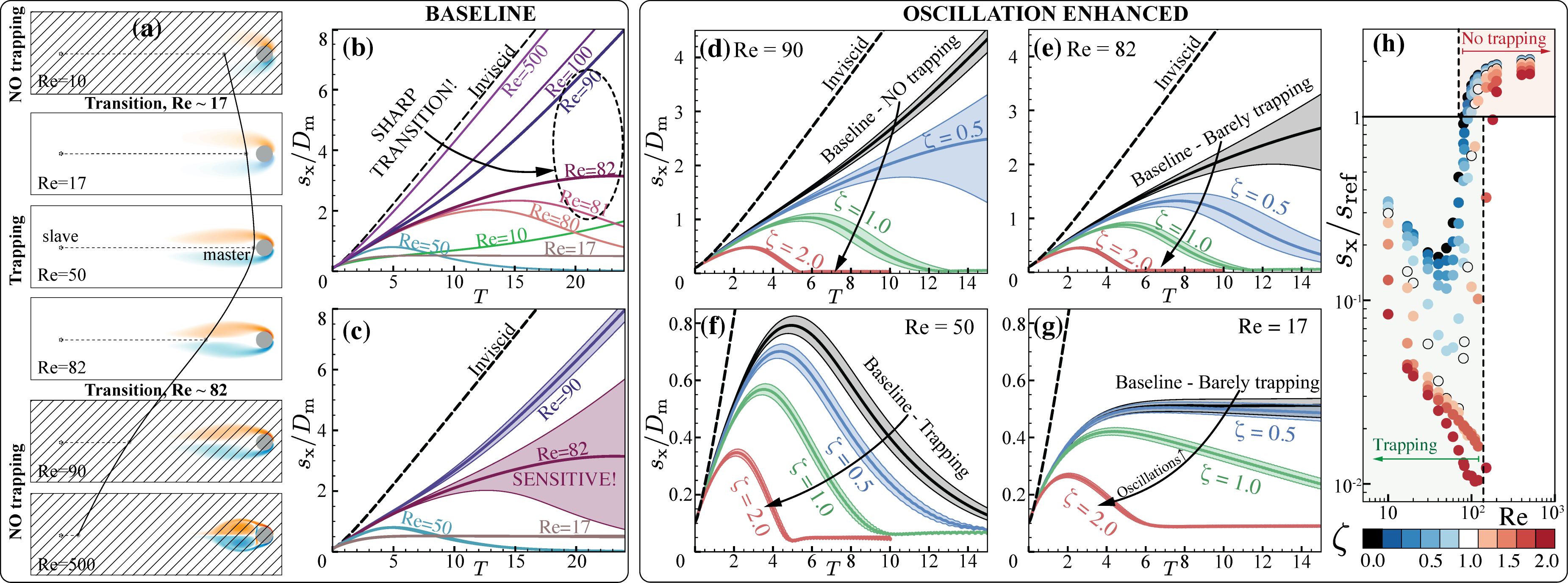}}
	\vspace{-12pt}
	\caption{Baseline cases: \textit{(a)} We observe a clear range of \( \Rey \) between which trapping and transport is achievable through linear motion. \textit{(b)} Plotting \( s_{\textrm{\scriptsize x}}(t)/D_{\textrm{\scriptsize m}} \) against \( T = 2U_{\textrm{\scriptsize l}} t/D_{\textrm{\scriptsize m}}\) reveals that transport is achieved between \( \Rey \approx 17\) and \(\Rey \approx 82 \), which then represent transitions between transport and non-transport regimes. At \(\Rey \approx 82 \), this transition is sharp (i.e. the system is unstable) and small changes in the \(\Rey\) lead to very different transport responses. \textit{(c)} Perturbing the initial slave location and observing \( s_{\textrm{\scriptsize x}}(t)\) reveals high sensitivity (shaded) of the system around \( \Rey \approx 82 \) (the top and bottom envelopes of the shaded region represent the characteristics for slaves with a \( \pm 2 \% \) perturbation to the initial separation distance). Oscillation enhanced cases: \textit{(d,e,f,g)} enabling oscillations enhances the ability of the system to transport the slave, across different \(\Rey \). \textit{(h)} We plot \( s_{\textrm{\scriptsize x}}/s_{\textrm{\scriptsize ref}} \textrm{ at } T=10\) for a number of \( (\Rey - \zeta )\) configurations to illustrate how transport--enhancement translates across a wider range of \( \Rey \).}
	\label{fig:main_panel}
	\vspace{-15pt}
\end{figure}

\vspace{-7pt}
We reproduce the results of the baseline cases of \cite{Gazzola:2012a} in \cref{fig:main_panel}a for clarity. As we increase \( \Rey \), we observe two distinct, sharp transitions (at \( \Rey \approx 17 \) and \( \Rey \approx 82 \)), between which the slave gets trapped and transported by the master due to linear motion (and associated wake) alone. This is quantified in \cref{fig:main_panel}b across a range of \( \Rey \), where we plot the normalized master--slave distance \( s_{\textrm{\scriptsize x}}/D_{\textrm{\scriptsize m}} \) against the non-dimensional time \( T = 2U_{\textrm{\scriptsize l}} t/D_{\textrm{\scriptsize m}}\). In this plot, a plateauing \( s_{\textrm{\scriptsize x}}\) or \( s_{\textrm{\scriptsize x}} \to 0\) indicates transport and is characteristic of \( 17 \lessapprox \Rey \lessapprox 82 \). For cases with \( \Rey \lessapprox 17 \) and \( \Rey \gtrapprox 82 \), \( s_{\textrm{\scriptsize x}} \) increases with time indicating that the slave is left behind as the master moves forward and hence it is not transported. At \( \Rey \approx 17 \) and \( \Rey \approx 82 \), \( s_{\textrm{\scriptsize x}} \) plateaus with increasing \(T\), indicating that the slave is trapped at a fixed distance from the master (and thus travels with the same speed). These two \(\Rey\) then denote the boundaries of transition between trapping and non-trapping regimes. Moreover, at \( \Rey \approx 82 \), this transition is sharp, as small changes in \(\Rey\) (from \(80\) to \(82\) to \(90\)) lead to large changes in the transport characteristics. Drawing from prior works \citep{lin2018effects}, we expect this sharpness (or sensitivity) to also be reflected in changes of initial master--slave separation \( s_{\textrm{\scriptsize x}}(0)\). We probe this sensitivity by perturbing \( s_{\textrm{\scriptsize x}}(0)/D_{\textrm{\scriptsize m}}\) by \( \pm 2 \% \) and observing the resulting transport characteristics. We depict this for a few key \( \Rey\) in \cref{fig:main_panel}c, wherein the shaded regions highlight the deviation associated with the perturbations. As expected, the system response is seen to be very sensitive at the transitional \( \Rey \approx 82\), relative to larger and smaller \( \Rey \). We consider this sensitivity as an opportunity to enhance transport. Indeed a carefully constructed flow perturbation can `pull' the system from the non-transport regime into the sensitive region and make it `jump over', enabling transport.

We choose to control the system by perturbing the flow via viscous streaming, by oscillating the master at different levels of intensity \(\zeta = R_{\textrm{\scriptsize o}}/\Rey \), and report our observations in \cref{fig:main_panel} (d-g). For \(\Rey = 90 > 82\) (above transition, no baseline transport) shown in \cref{fig:main_panel}d, mild oscillations (\(\zeta = 0.5\), blue bands) assist transport and bring the system into the sensitive region. Increasing \(\zeta\) further pushes the system well beyond the transition, enabling (and quickening) slave transport (green and red bands). \Cref{fig:main_panel}(e-g) shows the oscillation-enabled characteristics for \(\Rey = 82, 50 \) and \(17\), where we consistently observe a similar behaviour. We then undertake a parametric investigation to further characterize the effect of superimposing oscillations, systematically spanning \(\Rey\) between \(10\)--\(500\) and \(\zeta\) between \(0\)--\(2\). We depict this in \cref{fig:main_panel}h, where we plot \( s_{\textrm{\scriptsize x}}/s_{\textrm{\scriptsize ref}}\) (at \(T = 10\)) against \(\Rey\), for different \(\zeta\). Here \(s_{\textrm{\scriptsize ref}}\) is the separation distance (at the same time \(T = 10\)) of the reference baseline at \(\Rey \approx 82\), i.e. at the transition between transport and non-transport simulations when oscillations are not active (i.e. \(\zeta = 0\)). This qualitatively means that cases in \cref{fig:main_panel}h with \(s_{\textrm{\scriptsize x}}/s_{\textrm{\scriptsize ref}} < 1\) transport the slave. We then observe that while oscillations always assist transport, higher \(\zeta\) values are necessary for trapping at higher \(\Rey\), and beyond \(\Rey \approx 200\) oscillations are no longer able to drive the system into the transport regime, and the slave is then left behind. The system still retains its inherent sensitivity to \(\Rey\), apparent from the sharp jump between trapping/non-trapping cases across \( 100 \lessapprox \Rey \lessapprox 200\), for any fixed \(\zeta\). We conclude that introducing oscillations, modulated by \(\zeta\), enhances inertial particle transport across a wide range of \(\Rey\). The causal mechanism, be it viscous streaming or other wake-oscillation interactions, is ascertained in the next sections.

\vspace{-15pt}
\subsection{Robustness}\label{ss:robustness}
\begin{figure}[h!]
	\centerline{\includegraphics[width=\linewidth]{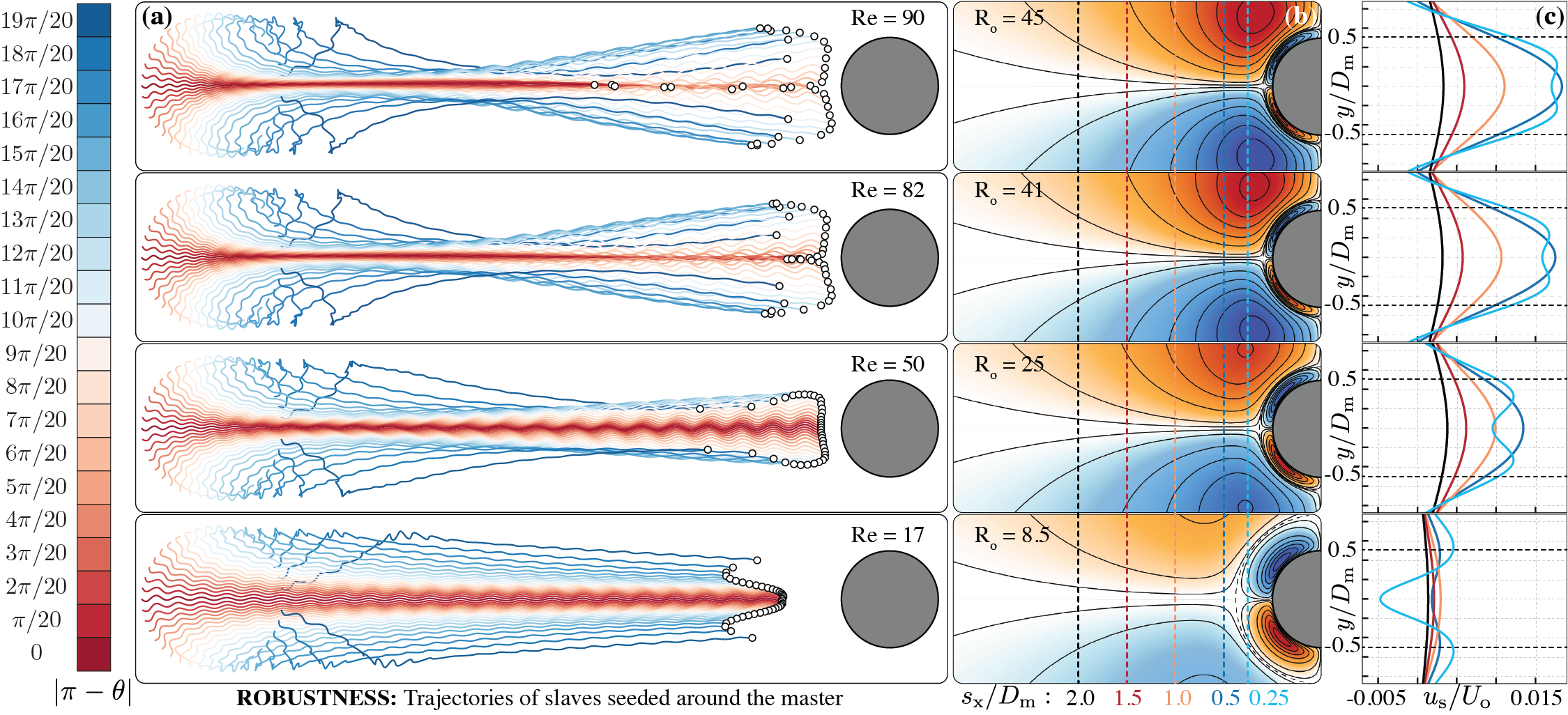}}
	\vspace{-12pt}
	\caption{Robustness: (\textit{a}) Seeding slaves azimuthally around the master (with \(\zeta = 0.5\)) and plotting their trajectories (till \(T=14\)) reveals that all slaves get transported for different \( \Rey \in [17, 82]\), lending to the robustness of this transport enhancement strategy. The trajectories can be correlated to (\textit{b}) the associated pure streaming fields (colored by streamfunction with the solid lines indicating streamlines). Dashed lines indicate different stations at which we plot (\textit{c}) streaming-induced velocities. We do not consider slaves in the comparison against pure streaming fields, with the implicit assumption that they do not drastically perturb the flow.}
	\label{fig:robustness_panel}
	\vspace{-15pt}
\end{figure}

\vspace{-7pt}
Given the sensitivity of the system (\cref{fig:main_panel}c) to initial horizontal slave separation, it is important to further characterize the spatial robustness of the proposed transport strategy. Increasing \(\zeta\) aids transport, but so far we tested only slaves initially located directly behind the master (\cref{fig:main_panel}h). We then initialize slaves (diameter \(D_{\textrm{\scriptsize s}} = \squart D_{\textrm{\scriptsize m}}\)) at the same surface-to-surface distance (\( 0.1D_{\textrm{\scriptsize m}} \)), but at different azimuthal positions \( \theta \) around the master, in separate simulations. We choose a representative \( \zeta = 0.5 \) and vary only the \( \Rey \). \Cref{fig:robustness_panel}a illustrates the slave trajectories at \( T = 14 \) for representative \( \Rey \), colored by their initial azimuthal position. We observe that almost all slaves get transported, irrespective of their initial azimuthal positions (and initial radial position perturbations, see Supplementary Material). However we notice differences between the slave trajectories at different \(\Rey \).

These differences can be comprehended by analyzing the corresponding pure streaming (\cref{fig:robustness_panel}b,c) flow fields. \Cref{fig:robustness_panel}b shows the time-averaged streamlines, where we also depict the stations at which the corresponding streaming velocities (\cref{fig:robustness_panel}c) are portrayed. At \( \Rey = 17\), a thicker DC layer (with reverse flow, blue line in \cref{fig:robustness_panel}c) `cushions' the master and prevents the slaves from attaining close proximity. All the other cases have a DC layer of almost constant thickness, thus lending to qualitative differences, highlighted by slaves closely surrounding the master's posterior. Comparing \( \Rey = 50\) and \( \Rey = 82\), we notice that in the latter slaves with initial positions \( | \pi - \theta | \gtrsim 15\pi/20 \) (\cref{fig:robustness_panel}a, colorbar) are transported further, due to more favourable upstream velocity profiles. Additionally, slaves almost directly behind the master (with initial positions \( | \pi - \theta | \lesssim \pi/5 \)) trail for \( \Rey = 90\) as compared to \( \Rey = 82\) and \(\Rey = 50\). This is explained by the fact that \( \Rey = 90\) is characterized by a baseline that cannot trap. Mild oscillations (\(\zeta = 0.5\)) are then just barely capable of `pulling' the slaves close enough to be trapped and transported. We conclude that oscillation-based transport strategy is robust overall. Moreover the resulting slave trajectories are found to be consistent with streaming-induced velocities, suggesting that this is the responsible transport enhancement mechanism at play.

\begin{figure}[h!]
	\centerline{\includegraphics[width=\linewidth]{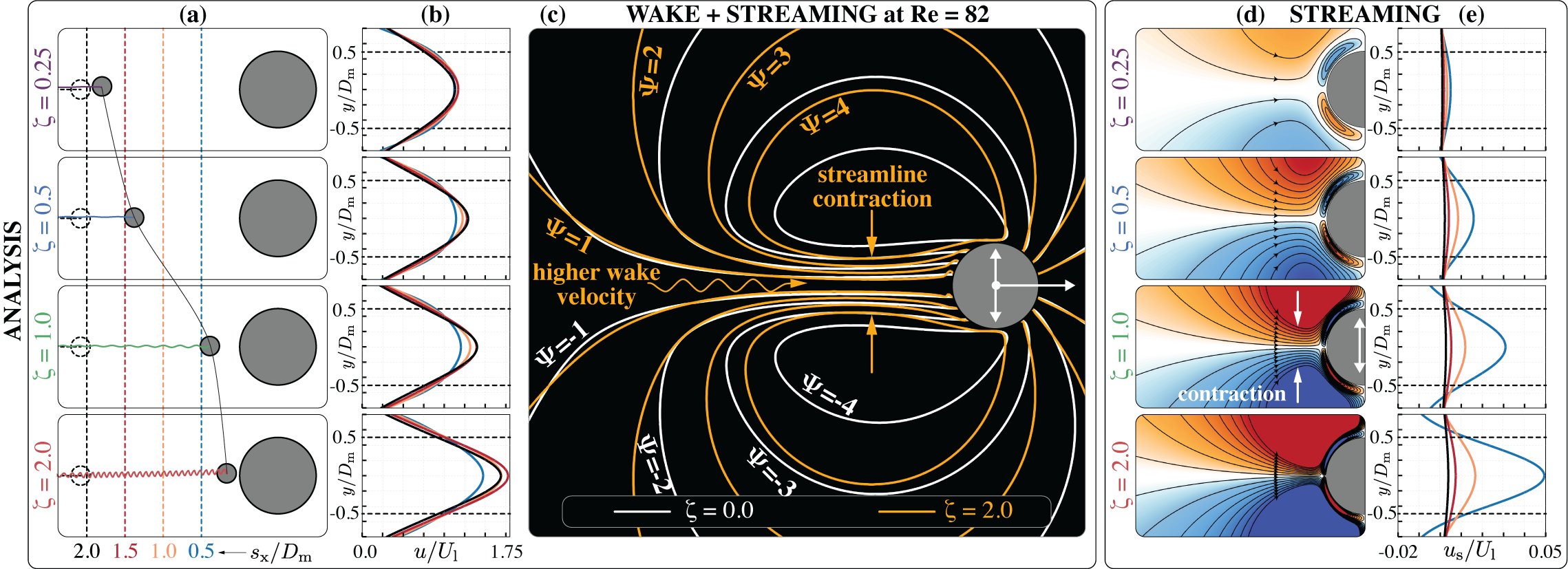}}
	\vspace{-10pt}
	\caption{Analysis: Fixing \(\Rey=82\) and increasing \(\zeta\) leads to (\textit{a}) better transport due to flow acceleration in the wake. The trajectories are drawn till \(T = 14\) with the dashed black lines representing the baseline case. (\textit{b}) Time-averaged velocity profiles (over 20 oscillation cycles) at the marked stations of (\textit{a}). The observed accelerations are due to streamline-contraction in the wake, visualized in (\textit{c}) for representative cases without (\(\zeta=0\), white streamlines) and with (\(\zeta=2\), orange streamlines) streaming. Corresponding streaming-only fields---(\textit{d}) streamlines and (\textit{e}) velocities at the same stations as (\textit{a,b})---explain the behaviour of (\textit{a,b,c}).}
	\label{fig:analysis_panel}
\end{figure}

\vspace{-15pt}
\subsection{Flow analysis at the transport/non-transport boundary}\label{ss:analysis}
\vspace{-7pt}
Here, we further focus on the fluid mechanisms at play, by drawing parallels between the system at the transitional \(\Rey = 82\) with different \( \zeta \) and corresponding pure streaming cases. \Cref{fig:analysis_panel}a pictures the slave trajectories for increasing \(\zeta\), highlighting improvement in transport. This is due to a corresponding increase in the wake velocities as seen in \cref{fig:analysis_panel}b, where we plot the time-averaged velocities, at different stations marked in \cref{fig:analysis_panel}a. The averaged streamlines (streamfunction iso-contours) of \cref{fig:analysis_panel}c for cases with (\(\zeta=2\)) and without (\(\zeta=0\)) oscillations explain this increase in wake velocity. Oscillations `tilt' the streamlines backwards, simultaneously compressing them directly behind the master, thus locally increasing upstream flow velocities. The degree of this contraction increases with \(\zeta\). This behaviour is consistent with the corresponding streaming-only patterns of \cref{fig:analysis_panel}d where larger oscillation intensities increasingly push the outer eddies towards the mid-plane, causing streamline contraction and increasing flow velocities, as further quantified in \cref{fig:analysis_panel}e. We thus identify streamline contraction as the primary cause for transport enhancement, with the viscous streaming mechanism driving this contraction.

\vspace{-15pt}
\subsection{Design}\label{ss:design}
\begin{figure}[h!]
	\centerline{\includegraphics[width=\linewidth]{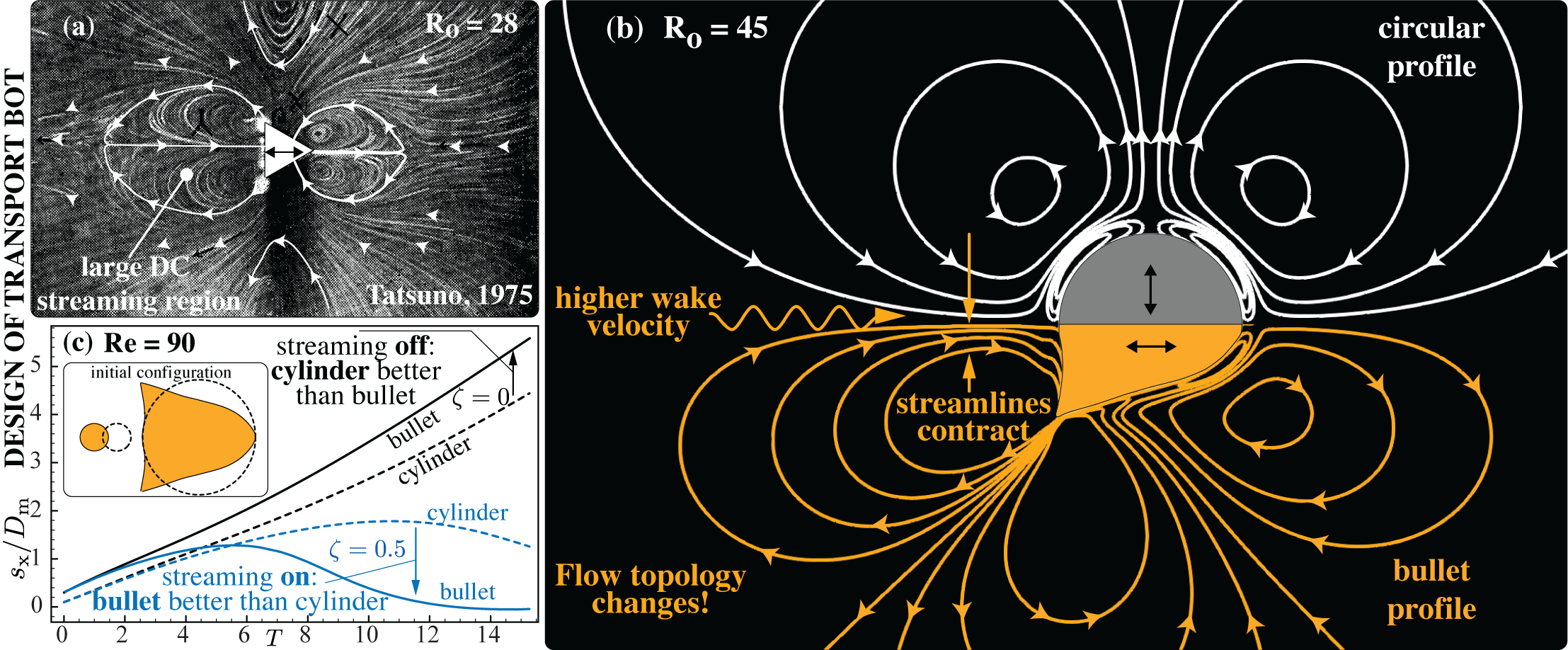}}
	\vspace{-12pt}
	\caption{Design: (\textit{a}) Inspired by streaming from triangles \citep{tatsuno1975circulatory}, we design the `bullet' shape of (\textit{b}) to induce a more favourable streaming field (more streamline contraction at \(R_{\textrm{\scriptsize o}} = 45 \)) which is reflected in better (\textit{c}) transport characteristics (at \(\Rey = 90\)) relative to a simple circular cylinder. This is in spite of our initial configuration (inset) that penalizes the bullet by placing the slave further away.}
	\label{fig:design_panel}
	\vspace{-7pt}
\end{figure}

\vspace{-7pt}
If streaming is indeed the responsible agent for transport enhancement, we should be able to design new geometries that produce more favourable streaming fields that actually translate in improved slave transport once tested in our setup of~\cref{fig:framework}b. We start by considering streaming only, and draw inspiration from the visual investigation of \citep{tatsuno1975circulatory} on streaming triangles. These are shown to produce a large DC recirculation region (\cref{fig:design_panel}a), which can be leveraged to trap and carry along passive cargoes. We then `borrow' two key components of this geometry---rear high-curvature tips and fore-aft symmetry breaking---to design a master with a `bullet' cross-section (\cref{fig:design_panel}c, inset---refer to Supplementary Material for shape parametrization).
This geometry is chosen to facilitate its comparison with the circular cylinder. Additionally, consistent with \citep{tatsuno1975circulatory} and differently from our calculations prior to this section, we change the direction of oscillation, from transverse (vertical) to longitudinal (horizontal), to produce the large posterior DC streaming layer observed in \cref{fig:design_panel}a. In \cref{fig:design_panel}b, we show that our new design constricts the streamlines further than a transversely-oscillating circle. This is reflected in its transport characteristics (\cref{fig:design_panel}c): while the circular cylinder performs better than the bullet when streaming is not active (no oscillations, pure master linear motion), the bullet outperforms its circular counterpart when streaming is enabled, even for mild oscillation intensities (\(\zeta = 0.5\)).

\begin{figure}[h!]
	\centerline{\includegraphics[width=\linewidth]{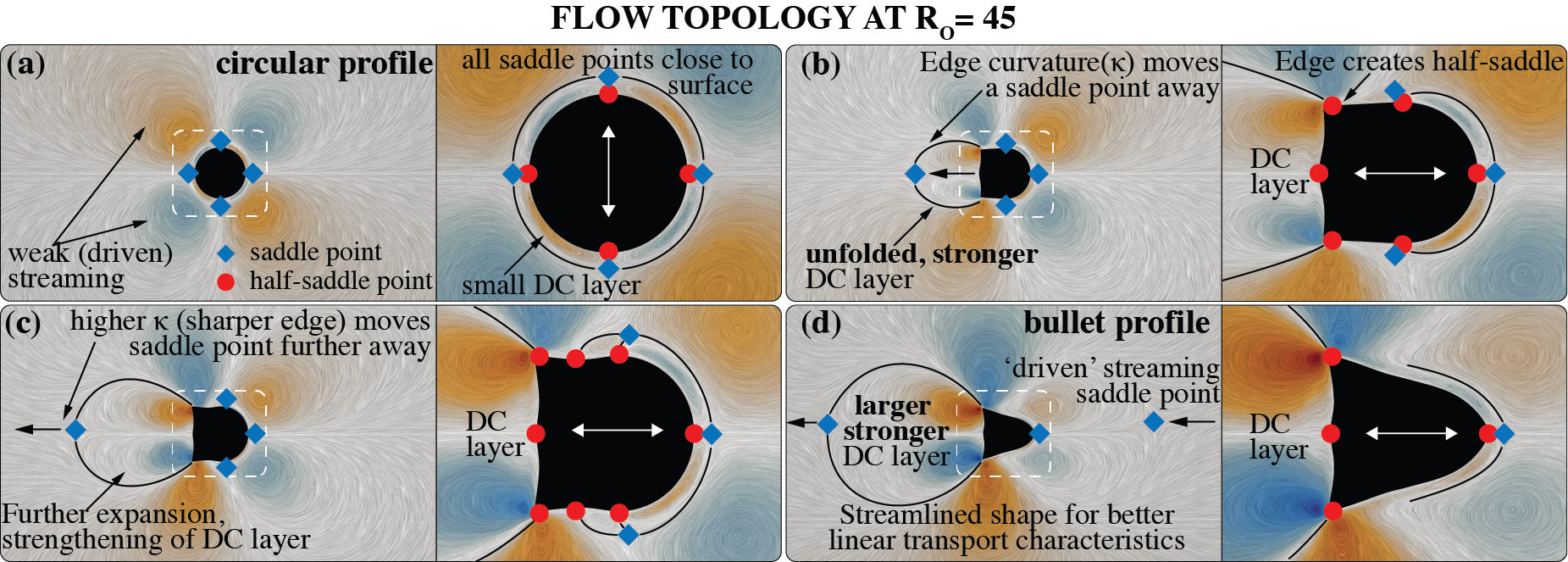}}
	\vspace{-10pt}
	\caption{Topology: We study the flow topological response to incremental shape changes (from the \textit{(a)} circle to the \textit{(d)} bullet) via Line Integral Convolution (LIC) of the streaming velocity field. This reveals a route for the rational design of streaming-based transport devices. The key idea involves introducing rear high-curvature points and fore-aft symmetry breaking to enlarge/strengthen DC layer streaming, the extent of which is indicated by the saddle points and marked by solid black lines. \textit{(b,c)} are intermediate shapes that illustrate the process.}
	\label{fig:topology_panel}
	\vspace{-15pt}
\end{figure}

The reason for the increased streamline contraction above lies in the flow topology change associated with the introduction of multiple curvatures and asymmetry in the master geometry (\cref{fig:design_panel}b). To understand how this topological transition occurs, we progressively morph (\cref{fig:topology_panel}) the circular cylinder into the bullet, and track the behaviour of critical points in the flow (these are saddle and half-saddle points \citep{hunt1978kinematical}, which are sparse yet complete representations of the global field \citep{perry1987description}). In the context of streaming, saddle points indicate the physical extent of the DC layer, marked by solid black lines in \cref{fig:topology_panel}. Morphing may create, destroy, merge or displace saddle and half-saddle points, which we leverage to manipulate the DC layer. Introducing high rear curvature points to the circle in \cref{fig:topology_panel}b creates two new half-saddles, which allows the rear saddle point to move away from the surface, thus `unfolding' the DC layer. The extent of this offset and the corresponding strength of the DC streaming region is related to the magnitude of the tip curvature. This is seen in \cref{fig:topology_panel}c where tip curvature increase enlarges the DC layer. Streamlining the master geometry further strengthens the DC layer and gives rise to our final bullet (\cref{fig:design_panel}d) design. Thus by manipulating the shape (and the oscillation direction) of the master---which dictates the flow topological response---we can rationally design configurations that improve slave transport.

We conclude this section wherein we have conducted extensive two-dimensional investigations to analyse, understand and design streaming-based transport systems. In the next section, we carry forth these ideas to three-dimensions.

\section{Transport in three dimensions}\label{sec:3d}
\begin{figure}[h!]
	\begin{center}
		\centerline{\includegraphics[width=\linewidth]{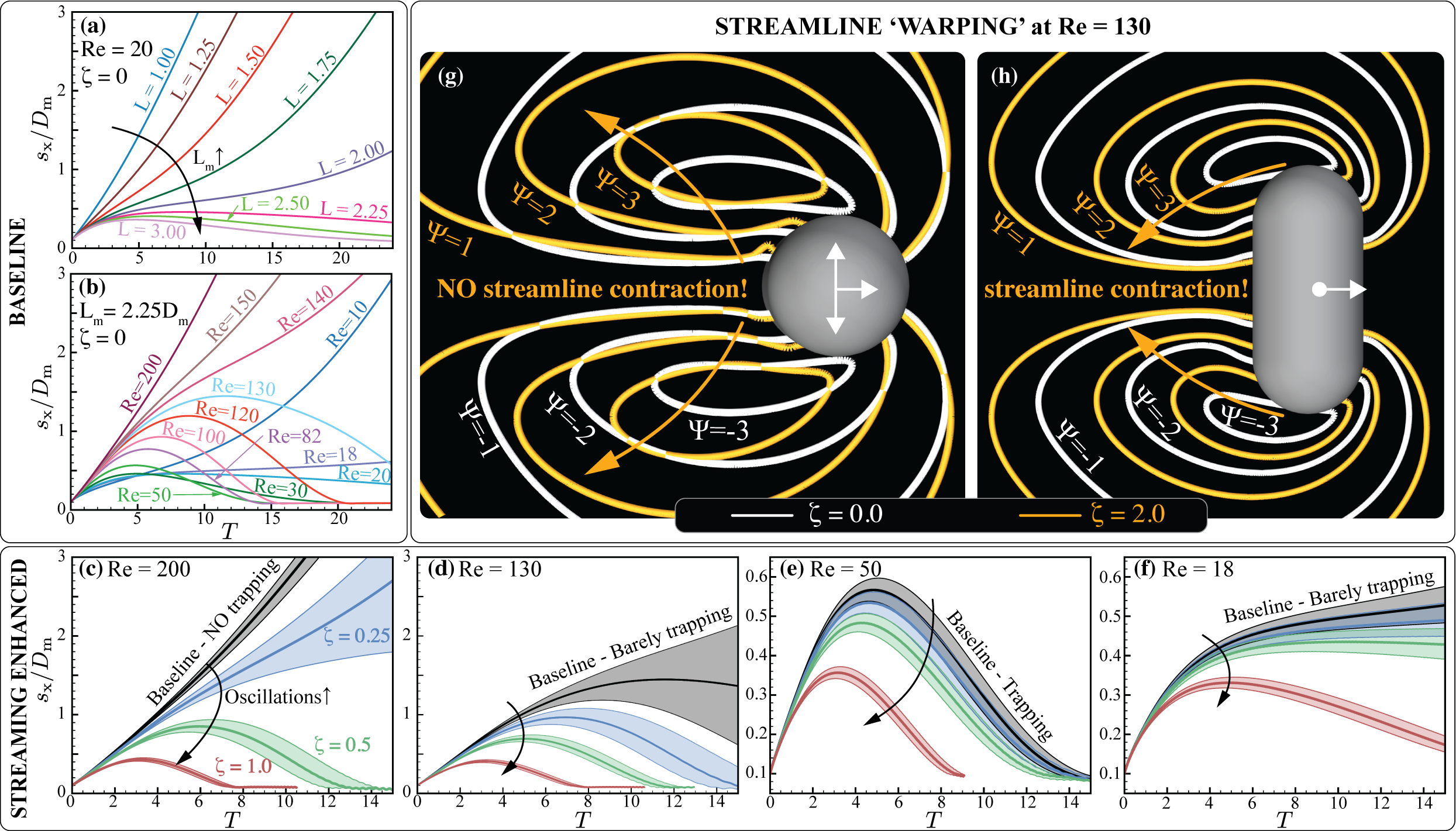}}
		\vspace{-10pt}
		\caption{Baseline: \textit{(a)} Increasing \(L_{\textrm{\scriptsize m}}\) suggests trapping via linear motion is feasible at \( \Rey = 20\) \textit{(b)} Performance of \(L_{\textrm{\scriptsize m}}=2.25D_{\textrm{\scriptsize m}}\) across different \( \Rey \) reveals transport/non-transport transitions at \( \Rey \approx 18\), \(\Rey \approx 130 \) (barely trapping and sharp transition). Streaming enhanced: \textit{(c,d,e,f)} Enabling oscillations enhances the ability of the system to trap/transport the slave across different \( \Rey \). Comparing the streamlines for cases without (\(\zeta=0\)) and with (\(\zeta=2\)) oscillation at \(\Rey = 130\) reveals that streamline contraction does not happen on the \textit{(g)} oscillation plane but on the \textit{(h)} plane perpendicular to the oscillation plane.
		}
		\label{fig:3D_results}
	\end{center}
	\vspace{-25pt}
\end{figure}

\vspace{-10pt}
Here we investigate how streaming enhancement strategies translate in a three-dimensional transport setting. The use of spheres instead of cylinders does not lead to slave transport, in contrast with two-dimensional results (Supplementary Material). We then proceed to elongate the master body with an end-to-end length \(L_{\textrm{\scriptsize m}}\), producing a pill-shaped cylinder with hemispherical ends, striking a compromise between a sphere and a circular cylinder. \Cref{fig:3D_results}a demonstrates the master's ability to transport/trap the slave at \(\Rey = 20\) (supplementary material) via linear motion by increasing \(L_{\textrm{\scriptsize m}}\).
Evaluating the performance of the master with \(L_{\textrm{\scriptsize m}} = 2.25 D_{\textrm{\scriptsize m}}\) (sensitive, barely trapping regime) across different \(\Rey\) (\cref{fig:3D_results}b) reveals two transition regimes (at \( \Rey \approx 18 \) and \( \Rey \approx 130 \)), consistent with observations in two dimensions.

We again introduce oscillations and perturbations to a few key $Re$ (\cref{fig:3D_results}c--f) and observe the enhancement in transport capability of the system. We relate this enhancement to the streamline contraction ensued from streaming. However, contrary to the two dimensional counterpart, the contraction manifests on the plane perpendicular to the oscillation plane (\cref{fig:3D_results}g--h). While the concept of leveraging streaming for transport extends from two to three dimensions, the same contraction mechanism is found to be activated differently. A full three-dimensional characterization is beyond the scope of this study and will be pursued in a follow-up investigation.

\section{Conclusions}\label{sec:conclusion}
\vspace{-10pt}
We have shown that oscillations can be utilized to robustly improve transport in an idealized two-dimensional master--slave setting across \(1 \le \Rey \le 100\). The analysis of flow features identifies viscous streaming as the catalyst for this improvement. Leveraging this information, we designed geometries exhibiting more favourable streaming patterns, which resulted in improved slave transport. To that extent, we demonstrated a rational design approach by modifying the classic circular cylinder via the introduction of multiple curvatures and fore-aft symmetry breaking. Moreover, we showed that similar concepts extend to three-dimensions even though favourable streaming effects are activated differently.

In conclusion, we have highlighted viscous streaming as a robust mechanism for achieving flow and passive particle control in a regime associated with emerging miniaturized robotic applications, such as, for example, drug delivery. However, despite the potential of viscous streaming in the context of complex, moving geometries, we still understand surprisingly little. What is the role of body shape and curvature in streaming? How can streaming from multiple bodies be effectively used at a collective level? How do three-dimensional effects affect it? While we have presented a rigorous, extensive study in a simplified setting, all these questions remain avenues of future work.

\vspace{-10pt}
\section{Acknowledgements}
\vspace{-10pt}
We thank Sascha Hilgenfeldt for helpful discussions over the course of this work and Wim M. van Rees for technical support on three-dimensional simulations. We also thank the Blue Waters project (OCI-0725070, ACI-1238993), a joint effort of the University of Illinois at Urbana-Champaign and its National Center for Supercomputing Applications, for partial support.

\vspace{-10pt}
\section{Appendix}
\vspace{-10pt}

\appendix

\section{Two dimensional viscous flow mediated interactions--Numerics}
\vspace{-10pt}
We use the remeshed Vortex Method (rVM) algorithm described in \citep{Gazzola:2011a} to perform two and three dimensional viscous flow-structure interaction simulations. Here, we list our simulation methodology and parameters for reproducibility and completeness.

We simulate the master (diameter $ D_{\textrm{\scriptsize m}} =  0.111$m) and the slave (diameter $ D_{\textrm{\scriptsize s}} =  0.028$m) in a domain of physical size $ [0,1]^2 $ m$^2$, with constant grid size of $ 2048 \times 2048 $ in the $ x $ and $ y $ directions. The master is initialized at $ \boldsymbol{x}^0_{\textrm{\scriptsize m}} = [0.104, 0.5] $ m and moves laterally with a constant velocity of $ U_{\textrm{\scriptsize l}} = |D_{\textrm{\scriptsize m}}|$ ms$ ^{-1} $. The slave is initialized at a surface-to-surface distance of $ 0.1D_{\textrm{\scriptsize m}} $ m behind the master at $ \boldsymbol{x}^0_{\textrm{\scriptsize s}} = [0.023, 0.5] $ m. The choice of the diameters and initial locations for the cylinders is made to maximize the resolution across both master and slave for a given simulation and for the non-dimensionless end time $ T = 2U_{\textrm{\scriptsize l}t}/D_{\textrm{\scriptsize m}} = 15 $. The number of grid points across the diameter of the master and slave are kept constant at $ 226 $ and $ 56 $ respectively. The viscosity $ \nu $ is set based on the linear Reynolds number $ \Rey = U_{\textrm{\scriptsize l}} D_{\textrm{\scriptsize m}}/\nu $. We oscillate the master with a fixed amplitude $ A = \thalf \epsilon D_{\textrm{\scriptsize m}}$, while the angular frequency $ \omega $ is determined through the oscillatory Reynolds number $ R_{\textrm{\scriptsize o}}  = \zeta \Rey = \epsilon \omega D_{\textrm{\scriptsize m}}/ (2\nu) $. We note that that both cylinders are density matched, with $ \rho  = 1$ kgm$^{-3}$.

\vspace{-15pt}
\section{Two dimensional inviscid flow mediated interactions--Algorithm}
\vspace{-10pt}
We discuss the numerical strategy used for the potential flow simulations and validate it against two benchmark problems. We once again consider two dimensional incompressible flow in an unbounded domain ($\Sigma$), in which $ n $ moving rigid bodies are immersed. We denote with $\Omega_i$ \& $\partial\Omega_i$, $i=1,\cdots n$ the support and boundaries of the solids, which are assumed to be of the same density of the fluid ($\rho_i=\rho=1$ kgm$^{-3}$). We further define $ \partial \Omega := \bigcup\limits_{i=1}^{n} \partial \Omega_{i} $. The algorithm is the same as \cite{munnier2010locomotion}, with minor modifications to simulate our master--slave systems.

\vspace{-15pt}
\subsubsection{Governing equations}
\vspace{-10pt}
The governing equations for the fluid in the inviscid limit are the Euler equations coupled with the incompressibility  constraint:
\begin{equation}
\frac{\partial \boldsymbol{u}}{\partial t} + \left( \boldsymbol{u} \bcdot \bnabla \right)\boldsymbol{u} = -\frac{1}{\rho}\nabla P~~~~\boldsymbol{x}\in\Sigma\setminus\Omega \label{eq:app_pot01}
\end{equation}
\begin{equation}
\bnabla \bcdot \boldsymbol{u}=0~~~~\boldsymbol{x}\in\Sigma\setminus\Omega \label{eq:app_pot02}
\end{equation}

The rigid solid body dynamics can be obtained by solving the Newton's equations of motion concurrently:
\begin{equation}
m_i \ddot{\boldsymbol{x}}_i = \boldsymbol{F}^H_i~~~~\label{eq:app_pot03}\\
\end{equation}
\begin{equation}
\frac{d(I_i\dot{\theta}_i)}{d t}=\boldsymbol{M}^H_i~~~~\label{eq:app_pot04}
\end{equation}

The causal force and moments in \cref{eq:app_pot03,eq:app_pot04} on the body result from the boundary conditions that couple the fluid--solid dynamics:
\begin{equation}
\boldsymbol{u} \bcdot \boldsymbol{n}(\boldsymbol{x}) = \boldsymbol{u}_i \bcdot \boldsymbol{n}(\boldsymbol{x})~~~~\boldsymbol{x}\in\partial\Omega_i \label{eq:app_pot05}\\
\end{equation}
where $ \boldsymbol{n} $ is the unit normal pointing towards the fluid and $ \boldsymbol{u_i} $ is the velocity of the rigid body $ i $, respectively. This encodes the no-through flow boundary condition.

To solve the above problem numerically we couch the solids and fluid systems into a larger one and solve for the total dynamics. This bypasses the need to calculate pressure and surface forces on the body as they are internal forces in the bigger system. As shown in \citep{lamb1932hydrodynamics}, one can adopt a Lagrangian perspective and solve for the total dynamics using the principle of minimal work (resulting in Euler--Lagrange equations). The dynamics (an initial value problem) then evolve in time according to the coupling between the solids and fluid (a boundary value problem). For the purpose of exposition we focus first on this coupling and then on evolving the system in time.

The idea is to simplify the non-linear problem \cref{eq:app_pot01} in the absence of initial ($ t = 0 $) vorticity using Helmholtz's theorem---which guarantees that vorticity remains absent in the flow at all times $ t > 0 $. We can then represent the conservative velocity vector as the gradient of a scalar potential function $ \phi (\boldsymbol{x})$, i.e. $  \nabla \phi(\boldsymbol{x}) = \boldsymbol{u}(\boldsymbol{x})$. As the velocity vector is always solenoidal (from \cref{eq:app_pot02})---the problem of solving \cref{eq:app_pot01,eq:app_pot02} is cast to equivalently solving the following Laplace equation.
\begin{equation}
\nabla^2 \phi (\boldsymbol{x})= 0 ~~~~\boldsymbol{x}\in\Sigma\setminus\Omega    \label{eq:app_pot06}
\end{equation}
with the boundary conditions
\begin{equation}
\nabla\phi (\boldsymbol{x}) \bcdot \boldsymbol{n}(\boldsymbol{x}) = (\boldsymbol{u}_i + \dot{\theta}_{i} \times (\boldsymbol{x} - \boldsymbol{x}_{i}^{COR} ) \bcdot \boldsymbol{n} (\boldsymbol{x}) ,~~~~\boldsymbol{x}\in\partial\Omega_i \label{eq:app_pot07}\\
\end{equation}
\begin{equation}
\nabla \phi (\boldsymbol{x}) = 0 \label{eq:app_pot08}, ~~~~\boldsymbol{x} \to \infty
\end{equation}
where $\dot{\theta}_{i}, \boldsymbol{x}_{i}^{COR}$ represent the angular velocity and the center of the rigid body $ i $. We thus solve an exterior Neumann Boundary Value Problem (NBVP), with \cref{eq:app_pot08} necessitating the flow decay at large distances. Following \citep{lamb1932hydrodynamics}, we decompose the potential field $ \phi (\boldsymbol{x})$ into elementary Kirchhoff potentials  $ \mathcal{X}_i(\boldsymbol{x}) $, $\boldsymbol{\varphi_i}(\boldsymbol{x})$ such that
\begin{equation}
\phi(\boldsymbol{x}) = \sum_{i=1}^{n}(\dot{\theta}_{i} \cdot \mathcal{X}_i(\boldsymbol{x}) + \boldsymbol{u}_i \bcdot \boldsymbol{\varphi_i}(\boldsymbol{x})) \label{eq:app_pot09}
\end{equation}
While this linear decomposition increases the algorithmic complexity of the problem, it enables us to separate the individual contribution from every rigid body velocity component in the flow. This is useful in calculating the added mass contributions resulting from the adjacent fluid being accelerated by the motion of the immersed solid bodies. The inertia (or) the mass matrix for the system $ \mathsfbi{M} $ thus consists of the solid body inertia and added mass (fluid) contributions. $ \mathsfbi{M} $ is a block matrix with the blocks $ \mathsfbi{M}_{ij} $ given by
\begin{equation}
\label{eq:addedmass_app_pot10}
\setlength{\arraycolsep}{4pt}
\renewcommand{\arraystretch}{1.3}
\mathsfbi{M}_{ij} = \left[
\begin{array}{ccc}
\int_{\Sigma \setminus \Omega} \nabla \mathcal{X}_i \bcdot \nabla \mathcal{X}_j & \int_{\Sigma \setminus \Omega} \nabla \mathcal{X}_i \bcdot \nabla \varphi^1_j & \int_{\Sigma \setminus \Omega} \nabla \mathcal{X}_i \bcdot \nabla \varphi^2_j\\
\int_{\Sigma \setminus \Omega} \nabla \varphi^1_i \bcdot \nabla \mathcal{X}_j & \int_{\Sigma \setminus \Omega} \nabla \varphi^1_i \bcdot \nabla \varphi^1_j & \int_{\Sigma \setminus \Omega} \nabla \varphi^1_i \bcdot \nabla \varphi^2_j\\
\int_{\Sigma \setminus \Omega} \nabla \varphi^2_i \bcdot \nabla \mathcal{X}_j & \int_{\Sigma \setminus \Omega} \nabla \varphi^2_i \bcdot \nabla \varphi^1_j & \int_{\Sigma \setminus \Omega} \nabla \varphi^2_i \bcdot \nabla \varphi^2_j\\
\end{array}  \right]
+
\left[
\begin{array}{ccc}
I_j \delta_{ij} & 0 & 0\\
0 & m_j \delta_{ij} & 0\\
0 & 0 & m_j \delta_{ij}\\
\end{array}  \right]
\end{equation}
where $ i, j  \in \{ 1, \cdots, n \}$ represent the $ i^{th} $ and $ j^{th} $body contributions, $ \boldsymbol{\varphi}$ has $[\varphi^1, \varphi^2] $ as its components, $ \delta_{ij} $ represents the Kronecker-delta function and the $  \boldsymbol{x} $ dependence on the integrands is implicit. This block captures the total finite inertia resulting from the presence and motion of both the moving bodies $i , j$ and the fluid surrounding them, thus rendering it important for the collective dynamics of the system.

This dynamics evolve according to the Euler--Lagrange formula, where we consider the total system energy functional as the Lagrangian function to minimize. The kinematic energy of the $ i^{th} $ solid body is $ K_i = \thalf m_i |\boldsymbol{u}_i|^2 + \thalf I_i \dot{\theta}_{i}^2$. The total kinematic energy of the fluid is $ K_f = \thalf \rho \int_{\Sigma \setminus \Omega_i} |\nabla\phi|^2 (\boldsymbol{x}) d\boldsymbol{x}$. In this work we do not consider conservative barotropic forces and so the potential energy contribution is identically zero. The total system energy functional is thus $ L(\boldsymbol{q}, \dot{\boldsymbol{q}}) = K_f + K_1 + K_2 = \thalf \dot{\boldsymbol{q}}^T \mathsfbi{M}(\boldsymbol{q}) \dot{\boldsymbol{q}}$---a function of the state $ \boldsymbol{q} $ and its derivative $ \dot{\boldsymbol{q}} $. Here $ \boldsymbol{q} $ represents the degrees of freedom for the system i.e. the angular and Cartesian positions of all the bodies (i.e. $ \boldsymbol{q}  = [\boldsymbol{q}_1, \cdots , \boldsymbol{q}_n]$ with $\boldsymbol{q}_i = [\theta_{i} \quad x^1_{i} \quad x^2_{i}], i\in \{ 1, \cdots, n \}$).  Using this Lagrangian function $ L $ we derive the Euler Lagrange equation for the state $ \boldsymbol{q} $
\begin{equation}
\label{eq:eulerlagrange_app_pot11}
\frac{d}{dt}\frac{\partial L}{\partial \dot{\boldsymbol{q}}} - \frac{\partial L}{\partial \boldsymbol{q}} = 0
\end{equation}
which equivalently results in
\begin{equation}
\label{eq:disc_eulerlagrange__app_pot12}
\mathsfbi{M} \ddot{\boldsymbol{q}} + \langle \mathsfbi{\Gamma}(\boldsymbol{q}), \dot{\boldsymbol{q}}, \dot{\boldsymbol{q}} \rangle = 0
\end{equation}
where $ \mathsfbi{\Gamma}(\boldsymbol{q})$ is a rank-$ 3 $ tensor identified as the Christoffel symbol \citep{munnier2010locomotion} and the $ \langle \mathsfbi{\Gamma}(\boldsymbol{q}), \dot{\boldsymbol{q}}, \dot{\boldsymbol{q}} \rangle $ is shorthand for $ \mathsfbi{\Gamma}(\boldsymbol{q})^{k}_{ij} \dot{{q}}_{j} \dot{{q}}_{k} $. If $ M_{ij} $ denotes the $ (i,j) $ entry of $ \mathsfbi{M}$ $(i,j \in \{1, \cdots, 3n\})$ and $ q_i $ denotes the entries of $ \boldsymbol{q} $ $(i \in \{ 1, \cdots, 3n \} )$, then we define the Christoffel symbol $ \Gamma^{k}_{ij} $ by
\begin{equation}
\label{eq:christofell__app_pot13}
\Gamma^{k}_{ij} = \frac{1}{2} \left(	\frac{\partial M_{ki}}{\partial q_j} + \frac{\partial M_{kj}}{\partial q_i} - \frac{\partial M_{ij}}{\partial q_k} \right)
\end{equation}
These `shape derivative' $ \frac{\partial M}{\partial q} $ terms are calculated efficiently according to the formulation in \citep{munnier2010locomotion}. With all the above manipulations that follow from \cite{lamb1932hydrodynamics,Nair:2007,munnier2010locomotion}, we have reduced the governing nonlinear PDEs \cref{eq:app_pot01,eq:app_pot02} to a system of nonlinear ODEs \cref{eq:disc_eulerlagrange__app_pot12} which can be integrated efficiently.

\vspace{-15pt}
\subsubsection{Representation}
\vspace{-10pt}
To solve the NBVP \cref{eq:app_pot06,,eq:app_pot07,eq:app_pot08} at every timestep, we use Boundary Element Methods (BEMs) based on integral formulations of the Laplace equation. BEMs for the Laplace equation only need to be discretized on the surface---making them fast and efficient---thus eluding the problem of remeshing at every time step. The conversion of \cref{eq:app_pot06,eq:app_pot07,eq:app_pot08} to a formulation convenient for BEM is carried out by using Green's theorem, that reduces all volume integrals in \cref{eq:addedmass_app_pot10} to surface integrals:
\begin{equation}
\label{eq:greens_app_potnum_02}
\int_{\Sigma \setminus \Omega} \nabla \mathcal{T}_i \bcdot \nabla \mathcal{T}_j d\boldsymbol{x} = \thalf \int_{\partial \Omega_j} \mathcal{T}_{i} \partial_{\boldsymbol{n}} \mathcal{T}_{j} d\sigma_j + \thalf \int_{\partial \Omega_i} \mathcal{T}_{j} \partial_{\boldsymbol{n}} \mathcal{T}_{i} d\sigma_i
\end{equation}
where $ \mathcal{T} $ is a proxy for any of $ \mathcal{X}, \varphi^1, \varphi^2 $. Having transformed the volume Laplace problem to the equivalent boundary integral form, we realize that we only need the elementary Kirchhoff potentials on the boundaries $ \partial \Omega_i $ (that is the Dirichlet data), given its normal derivatives (the Neumann data, from \cref{eq:app_pot07}). We then represent, discretize and solve for $ \mathcal{T}$ on the boundaries only. The representation of elementary potentials of body $ i $ is done using finite terms (with cardinality $ m$) of a Fourier series on $ \partial \Omega_i $---the choice of the basis reflects the compact support and periodicity (with period $ 2 \pi $) of $ \partial \Omega_i $.

We obtain the Dirichlet data (and its tangential derivatives) on the boundary using the Neumann-to-Dirichlet operator \citep{atkinson1967numerical} for the 2D Laplace kernel $ G(\boldsymbol{x},\boldsymbol{y}) = \frac{1}{2 \pi} \int_{\partial \Omega_i} \log |\boldsymbol{x}-\boldsymbol{y}| d \sigma_{\boldsymbol{y}}$. This reads for $ i \in \{1, \cdots n \} $ and $ \boldsymbol{x} \in \partial \Omega $
\begin{equation}
\label{eq:n2d_app_potnum_03}
\mathcal{T}_i (\boldsymbol{x})- \frac{1}{\pi} \int_{\partial \Omega} \frac{(\boldsymbol{y} - \boldsymbol{x})}{|\boldsymbol{y}-\boldsymbol{x}|^2} \bcdot \boldsymbol{n} (\boldsymbol{y}) \mathcal{T}_i (\boldsymbol{y}) d \sigma_{\boldsymbol{y}} = -\frac{1}{\pi} \int_{\partial \Omega_i} \log |\boldsymbol{y}-\boldsymbol{x}| N_i (\boldsymbol{y}) d \sigma_{\boldsymbol{y}}
\end{equation}
where we have prescribed $ N_i (\boldsymbol{y})$---the Neumann data from \cref{eq:app_pot07}. The tangential derivatives of $ \mathcal{T}_i $ are necessary to calculate the shape derivatives. This is trivially done as once $ \mathcal{T}_i \in C^{\infty}$ is known, we can take its derivative in the tangential direction efficiently by using the spectral equivalent of the standard differentiation operator.

\vspace{-15pt}
\subsubsection{Discretization}
\vspace{-10pt}
To solve the integral equation \cref{eq:n2d_app_potnum_03}, we use Nystr\"{o}m discretization coupled with the (spectrally accurate) trapezoidal quadrature rule. To numerically evaluate the integrals, we split the integrand into singular and non-singular contributions and use the scheme suggested in \citep{atkinson1967numerical} to evaluate the former---the latter is trivial to integrate numerically. The interested reader is referred to \citep{atkinson1967numerical} for the theoretical and \citep{munnier2010locomotion} for the implementation details. We discretize the boundary $ \partial \omega_i $ by $ \upsilon_i = 2m_i + 1 $ points, where $ m_i $ is the finite number of Fourier modes represented on the boundary. We represent \cref{eq:n2d_app_potnum_03} in the discrete form by the equation $ \mathsfbi{A} \boldsymbol{t} = \boldsymbol{r} $, where $ \boldsymbol{t},  \boldsymbol{r} $ are the discrete equivalents of $ \mathcal{T} $ and the right hand side being solved for. We factorize $\mathsfbi{A}$ by standard LU-decomposition for the reasons listed in \citep{munnier2010locomotion}. Having evaluated the elementary potential on $ \partial \Omega_i $, we proceed to evaluate the mass matrix \cref{eq:addedmass_app_pot10} and the Christoffel symbols $ \mathsfbi{\Gamma} $ \cref{eq:christofell__app_pot13} at these boundaries. The acceleration in \cref{eq:disc_eulerlagrange__app_pot12} is then evaluated and the whole system can be marched forward in time.

We now deal with the time-marching scheme used in solving \cref{eq:disc_eulerlagrange__app_pot12}, which we rewrite as
\begin{equation}
\label{eq:timestep_app_potnum_01}
\frac{d}{dt}
\setlength{\arraycolsep}{4pt}
\renewcommand{\arraystretch}{1.3}
\left[
\begin{array}{c}
\dot{\boldsymbol{q}} \\
\boldsymbol{q}
\end{array}  \right]
=
\left[
\begin{array}{c}
- \mathsfbi{M}^{-1} \langle \mathsfbi{\Gamma}(\boldsymbol{q}), \dot{\boldsymbol{q}}, \dot{\boldsymbol{q}} \rangle \\
\dot{\boldsymbol{q}}
\end{array}  \right]
\end{equation}
As the boundaries of the solids $ \partial \Omega $ are assumed to be infinitely differentiable $ C^{\infty} $, parametrized with respect to a boundary tangent variable $ t \in [0, 2 \pi]$, we can infer that the RHS is then also $ C^{\infty} $. The above problem is then well posed and infinitely differentiable in time---making it a candidate for higher-order time stepping schemes. We use the \verb}LSODA} function from \verb}ODEPACK} that uses upto $ 13^{th} $ order accurate non-stiff (Adams) or stiff (BDF) method adaptively  based on the data. We fix the absolute and relative tolerances of our ODE solver to $ 1.49 \times 10^{-8} $, unless stated otherwise.

\vspace{-20pt}
\subsubsection{Validation}
\vspace{-10pt}
Here we consider two validation cases for our algorithm. The first benchmark case is the slave (radius $ b $) motion due to pure sinusoidal oscillations of the master (radius $ a $) in one direction, wherein we have a closed form governing ODE at large master--slave distances \citep{Nair:2007} for an inertialess master--slave pair. The slave transport is then purely due to the added-mass terms arising from the presence of the intermediate fluid. The nonlinear analytical ODE governing the slave position $ x_{\textrm{\scriptsize s}} $ for the case with $ a = b = \sqrt{2}^{-1} $m, $ \rho  = \pi^{-1}$ kg m$^{-3}$ and for pure sinusoidal oscillations of the master $ x_{\textrm{\scriptsize m}} = \sin(t) $ is
\begin{equation}
\label{eq:kanso_test}
\ddot{x}_{\textrm{\scriptsize s}} = -\frac{\sin{t}}{(x_{\textrm{\scriptsize s}} - \sin{t})^2} + \frac{2 \cos^2{t}}{(x_{\textrm{\scriptsize s}} - \sin{t})^3}
\end{equation}
\begin{figure}[h!]
	\centerline{\includegraphics[width=1.0\linewidth]{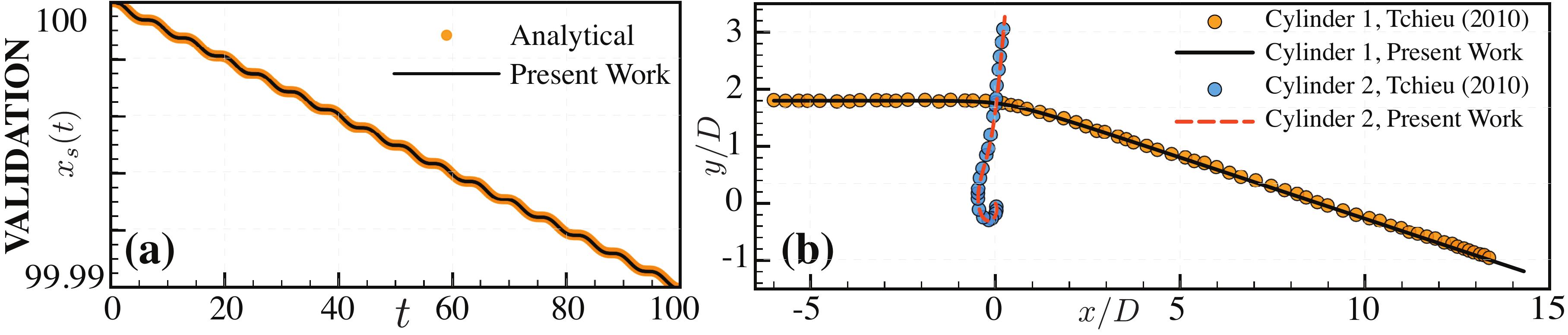}}
	\vspace{-10pt}
	\caption{Validation for the potential flow solver : (a) against a purely oscillating master--slave configuration and (b) against a near-collision event of two free cylinders}
	\label{fig:rifsi_validation}
	\vspace{-10pt}
\end{figure}

We now consider the same setup in our numerical solver using $ \upsilon_i = 121 $ points or $ m_i  = 60$ modes spatially on each cylinder (of density $ \rho = 1 \times 10^{-8} \sim 0$ each) and step forward in time with a constant $ \Delta t = 0.002 $s. The result from the solution of the closed form nonlinear ODE \cref{eq:kanso_test} and our numerical solution, for an impulsively started master, is shown in \cref{fig:rifsi_validation}(a). 

The second benchmark case is derived from \citep{Tchieu:2010a} and considers the near collision event of two `free/passive' cylinders in the flow. In this case, two neutrally buoyant ($ \rho_i = \rho = 1 $ kg m$^{-3}$) cylinders, of diameter $ 1 $ m each are initially placed in the cartesian plane at $ \boldsymbol{x}_1 = [-6.0 \quad 1.8]^T $ and $ \boldsymbol{x}_2 = [0.0 \quad 0.0]^T $ respectively. The former is given an initial velocity of $\dot{\boldsymbol{x}}_1 (0) = [1.0 \quad 0.0]^T$. It is noted that in the absence of the fluid both the cylinders will collide. The presence of the fluid acts as a `cushion', and helps prevent collision between the cylinders. The cylinders then nearly kiss one another---and any singular effects associated with the near-contact event need to be well resolved, making this a rigorous benchmark. We simulate this setup with our algorithm for $ \upsilon_i = 121 $ points each on the cylinders and with a constant $ \Delta t = 0.002 $s. The computed trajectory for both the cylinders is plotted in \cref{fig:rifsi_validation}(b) against the one in \citep{Tchieu:2010a}. The final velocity reported in \citep{Tchieu:2010a} for both the cylinders are  $\dot{\boldsymbol{x}}_1 (\infty) = [0.954 \quad -0.205]^T$ and $\dot{\boldsymbol{x}}_2 (\infty) = [0.030 \quad 0.216]^T$ upto three significant digits. With our solver we get $\dot{\boldsymbol{x}}_1 (\infty) = [0.954 \quad -0.205]^T$ and $\dot{\boldsymbol{x}}_2 (\infty) = [0.029 \quad 0.216]^T$ as the final velocities. Our results are thus in close agreement.

\vspace{-15pt}
\section{Two dimensional inviscid flow mediated interactions--Numerics}
\vspace{-10pt}
Using the above algorithm, we seek to replicate the two dimensional neutrally-buoyant master--slave configurations described in the main text in a potential flow context. For such simulations, we use a master cylinder of diameter $D_{\textrm{\scriptsize m}} = 1 $m and a slave of diameter $ D_{\textrm{\scriptsize s}} = 0.25 $m. The master is initially kept at $ \boldsymbol{q}_{\textrm{\scriptsize m}} = [0.0 \quad 0.0 \quad 0.0]^T $ and the slave is instantiated at $ \boldsymbol{q}_{\textrm{\scriptsize s}} = [0.0 \quad -0.725 \quad 0.0]^T $. The master is translated in the $ x $ direction and oscillated in the $ y $ direction. To avoid introducing impulse in the system (and thus eliminating bias) we ramp-up the motion of the master, as given by
\begin{equation}
\label{eq:potential_x_rampup}
x_{\textrm{\scriptsize m}} = U_{\textrm{\scriptsize l}} \left(t+\frac{1}{r}\ln\left(\frac{1+e^{r\left(c-t\right)}}{1+e^{rc}}\right)\right)
\end{equation}
\begin{equation}
\label{eq:potential_y_rampup}
y_{\textrm{\scriptsize m}} = \frac{\epsilon \omega D_{\textrm{\scriptsize m}} \sin\left(\omega t\right)}{2\left(1+e^{r\left(c-t\right)}\right)}
\end{equation}
where $ r = \frac{28\epsilon \omega}{3 \zeta}$ and $ c = \frac{15 \zeta}{14 \omega \epsilon}$ are ramping parameters. We set, for all cases, $ \epsilon = 0.05, \omega = 5 $ and choose $ U_{\textrm{\scriptsize l}} = \frac{U_{\textrm{\scriptsize o}}}{2 \zeta} = \frac{\epsilon \omega D_{\textrm{\scriptsize m}}}{4 \zeta} $. We then only vary the velocity ratio $ \zeta $ in our formulation in the simulations shown further below.

Unless stated otherise, we run all simulations with a constant time step $ \Delta t  = (200 \pi)^{-1} T $, where $ T = \frac{2 \pi}{\omega}$ is the time period of oscillation. We use $ \upsilon_i = 121 $ or $ m_i  = 60$ fourier modes to capture the master--slave interactions.

\vspace{-15pt}
\section{Master--slave baseline transport}
\vspace{-10pt}
Following~\citep{Gazzola:2012a}, we first understand the baseline transport before introducing transverse oscillations. We iterate again that the system was found to be sensitive in a viscous environment and so we first verify that our simulations can capture the comprehensive dynamics involved. We give detailed slave characteristics in the transitional $ \Rey $ in this section to highlight the importance of such transitions.

\vspace{-15pt}
\subsubsection{Validation against baseline for viscous flow}
\begin{figure}[h!]
	\centerline{\includegraphics[width=1.0\linewidth]{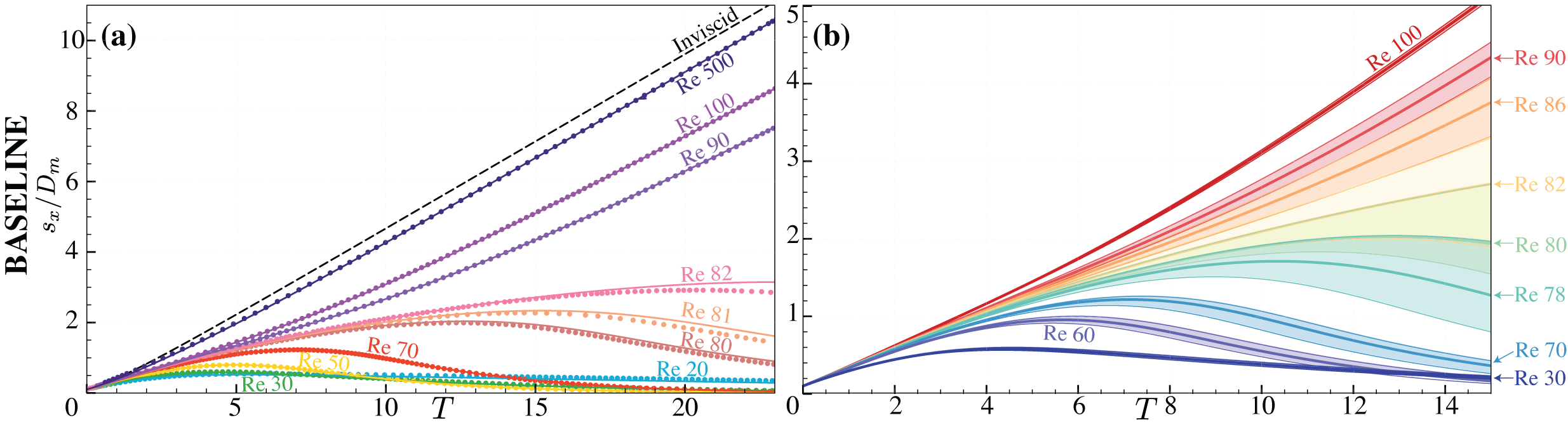}}
	\vspace{-10pt}
	\caption{Baseline transport: (a) Separation distance \( s_{\textrm{\scriptsize x}}/D_{\textrm{\scriptsize m}} \) versus non-dimensional time \( T = 2U_{\textrm{\scriptsize l}}t/D_{\textrm{\scriptsize m}}\) and (b) Sensitivity in the transition regime (from \(78 \lesssim \Rey \lesssim 86\) ). Notice that the maximum sensitivity occurs around \(\Rey \sim 82\). We give sensitivity at reference values of \(\Rey=30,60,70,90,100\) for comparison}
	\label{fig:chloe_match}
	\vspace{-15pt}
\end{figure}

\vspace{-10pt}
We first validate against the results of \citep{Gazzola:2012a}, where the slave response to the constant lateral motion of the master was investigated. We show this in \cref{fig:chloe_match}(a) for select values of $ \Rey $. Our simulation underpredicts transport at the transitional $ \Rey $, an unsurprising observation given the sensitivity of the flow to physical and simulation parameters. Apart from this, we obtain a good match for all the $ \Rey $ investigated. We also include the inviscid characteristics in this plot to assay pure potential effects, and see that our results correspond well with those of \citep{Gazzola:2012a}.
\vspace{-15pt}
\subsubsection{Transitional $\Rey$}
\vspace{-10pt}
We once again gauge the sensitivity of the baseline at the transitional \( \Rey \) below which slave's trapping and transport takes place given the baseline (\cref{fig:chloe_match}(a)). We perturb the initial conditions by the same amount (\(\pm 2 \% \)) as the main text and draw the envelopes of evolution of the perturbed slave trajectories for the transitional cases. From \cref{fig:chloe_match} we see that the cases within the transitional regime of ($ 78 \lesssim \Rey \lesssim 86 $) are especially sensitive. At \( \Rey = 81\) and \(82\), one trajectory traps while another does not and we denote them to be the Reynolds number of transition. Beyond this transitional regime, the baseline is insensitive to positional perturbations in the initial condition.

\section{Enhancing transport with streaming}
\vspace{-10pt}
\subsubsection{Estimating the magnitude of streaming-generated forces}
\vspace{-10pt}
We estimate that streaming generated forces for typical parameters are small, but not negligible---varying between \(0.1-10 \%\) of the wake-induced forces. An estimate of relevant forces per unit length induced by fluid flow past a static cylinder is given by \( F = \thalf C_{\textrm{\scriptsize D}}(\Rey) \rho u^2\), the coefficient \(C_{\textrm{\scriptsize D}}\) being a function of only \( \Rey \) \citep{panton2006incompressible}. Thus the force contribution from the wake is roughly \( F^{\textrm{\scriptsize w}} = \thalf C_{\textrm{\scriptsize D}}(\Rey) \rho U_{\textrm{\scriptsize l}}^2\) while streaming forces contribute forces on the order of \( F^{\textrm{\scriptsize s}} = \thalf C_{\textrm{\scriptsize D}}(R_{\textrm{\scriptsize s}}) \rho U_{\textrm{\scriptsize s}}^2\) (the implicit assumption of streaming flow being similar to a free-stream flow is justified as their time-scales are comparable). By definition, we have \( U_{\textrm{\scriptsize s}} = 2 \epsilon \zeta U_{\textrm{\scriptsize l}}\) and thus \( R_{\textrm{\scriptsize s}} = \epsilon \zeta \Rey \). Then the ratio of these forces couched in terms of \(\zeta\) and \(\Rey\) for a fixed \(\epsilon = 0.1\) is
 \begin{equation}
\label{eq:streaming_forces}
\frac{F^{\textrm{\scriptsize s}}}{F^{\textrm{\scriptsize w}}} \sim \frac{0.04 \zeta^2 C_{\textrm{\scriptsize D}}(0.2\zeta\Rey)}{C_{\textrm{\scriptsize D}}(\Rey)}
\end{equation}
We now consider only \(\Rey = 100\) for the purpose of exposition. With \(\zeta = 0.25\) and with \(C_{\textrm{\scriptsize D}}(100) \approx 1.1, C_{\textrm{\scriptsize D}}(5) \approx 3\) \citep{panton2006incompressible}, the ratio of forces is \( \sim 0.2 \% \). Considering a higher streaming intensity characterized by \(\zeta = 2\), we have \(C_{\textrm{\scriptsize D}}(40) \approx 1.2\) \citep{panton2006incompressible}, leading to a force ratio of \( \sim 17.4 \% \). The order of the streaming forces are thus suspected to be in the range of \(0.1\)--\(10 \% \), justifying their non-negligible contribution to the system dynamics.

\vspace{-15pt}
\subsubsection{Inviscid calculations: effect of oscillations and transitions}
\begin{figure}[h!]
	\centerline{\includegraphics[width=1.0\linewidth]{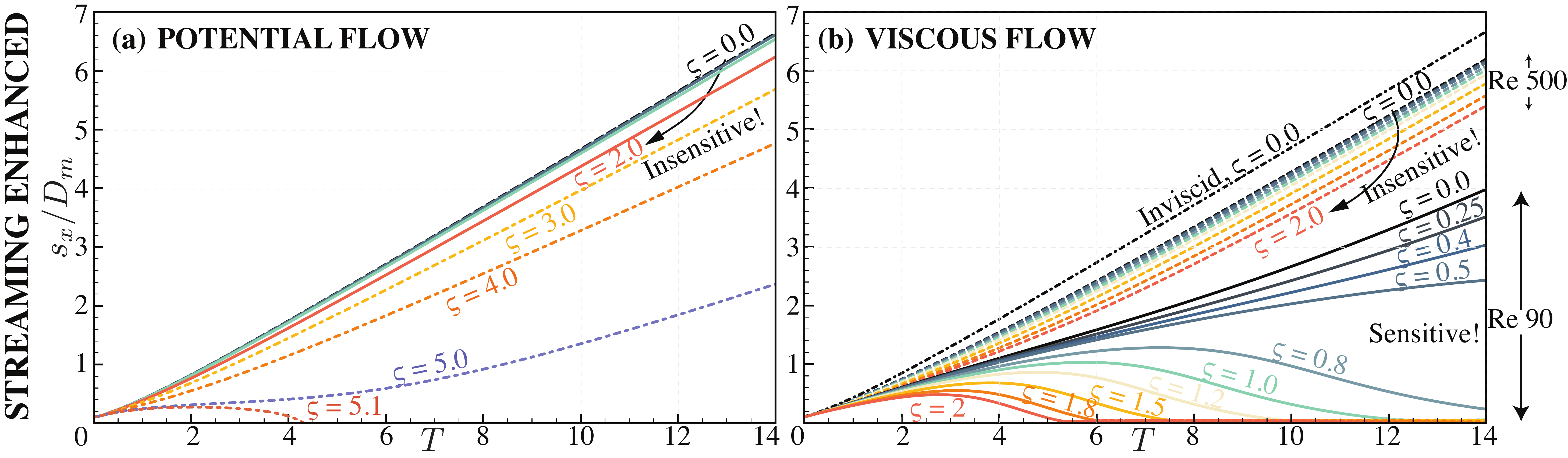}}
	\vspace{-10pt}
	\caption{Streaming enhanced transport: (a) Enhancement in transport from pure potential effects is minimal and high (\(\zeta \to 5\)) values are required for trapping, while enhancement in the (b) viscous case is appreciably more for finite \(\Rey\).}
	\label{fig:inv_visc_results}
	\vspace{-10pt}
\end{figure}

\vspace{-10pt}
We first investigate the transport characteristics arising from pure potential effects of the translation--oscillation strategy. We show these characteristics in \cref{fig:inv_visc_results}a, for various $ \zeta $ values. The system is initially insensitive to oscillations (and thus to perturbations in initial positions), while at high ($ >5 $) $\zeta$ we are able to trap and transport the slave due to potential effects alone. This shows that while the primary transport enhancement mechanism observed in our studies can not be attributed to inviscid interactions, these effects are non-negligible at higher $ \zeta $ values.

\subsubsection{Viscous calculations: effect of oscillations and transitions}
\vspace{-10pt}
A key result of ~\citep{Gazzola:2012a} was that enabling viscous effects assist transport in the case of a linearly translating master cylinder (our baseline). In addition to that, we find that the effect of oscillations in enhancing the ability of slave transport is amplified in a viscous regime. This is clearly seen when comparing \cref{fig:inv_visc_results}a and \cref{fig:inv_visc_results}b for some oscillation intensity (fixed \(\zeta\)). As seen in the main text, this enhancement can be attributed predominantly to the viscous streaming response. Using streaming, we perturb the slave's dynamics enough to fall in the transitional regime of \cref{fig:chloe_match}(b) at which point the first order linear motion of the master can trap the slave. This is highlighted in \cref{fig:inv_visc_results}b for the case of \(\Rey = 90\) at various \(\zeta\) values. We also call attention to the insensitivity in the characteristics at a high \(\Rey\) of \(500\), an unsurprising observation as we move towards the inviscid approximation.

\vspace{-15pt}
\subsubsection{Robustness to radial perturbations}
\begin{figure}[h!]
	\centerline{\includegraphics[width=0.8\linewidth]{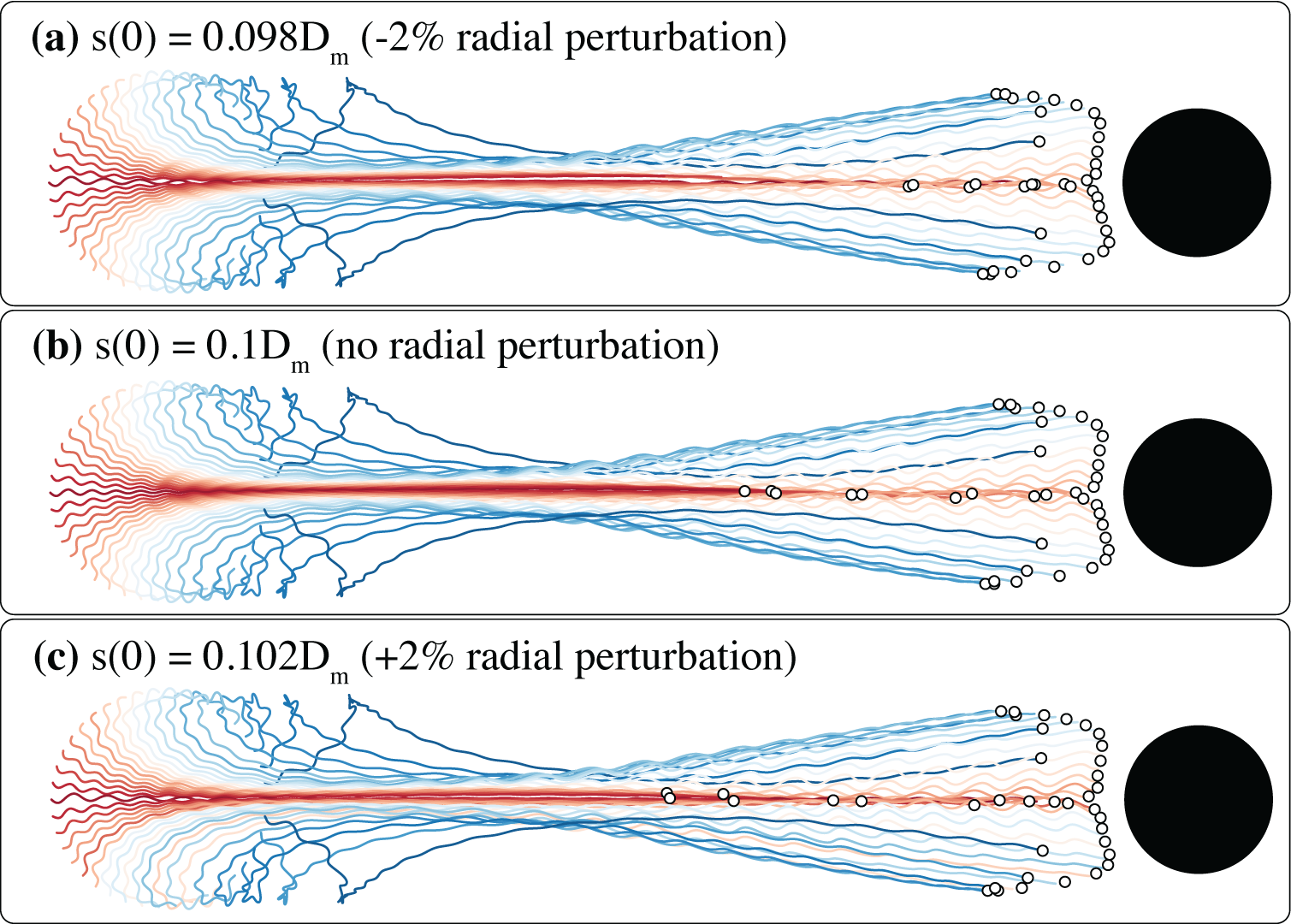}}
	\vspace{-10pt}
	\caption{Robustness of the streaming strategy to perturbations of the initial radial positions of seeded slaves}
	\label{fig:robustness_radial_perurb}
	\vspace{-10pt}
\end{figure}

\vspace{-10pt}
Streaming was shown to be a robust strategy with respect to the initial azimuthal positions of the slave. Here we consider the same setup, restricting the investigation to \( \Rey = 90, \zeta = 0.5\) while adding a \( \pm 2\% \) (radial) perturbation in the initial surface--surface distance between the master and the slave. The results are shown in \cref{fig:robustness_radial_perurb}. Streaming-based transport is also robust to radial perturbations. The only significant difference is seen in the trajectories of cases with \( | \pi - \theta | \lesssim \pi/5 \), due to the transitional nature of the baseline as stated in the main text.

\vspace{-15pt}
\subsubsection{Design: spline parametrization}
\begin{figure}[h!]
	\centerline{\includegraphics[width=0.4\linewidth]{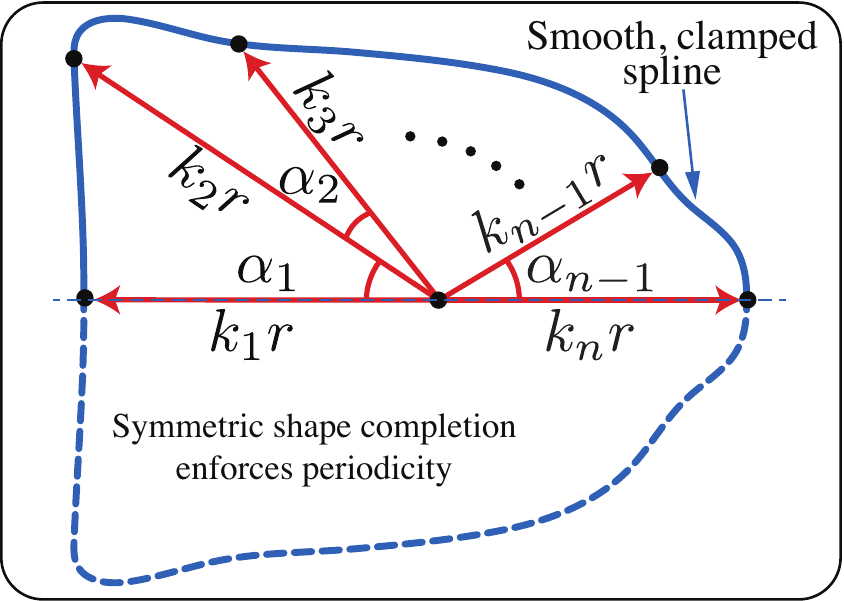}}
	\vspace{-10pt}
	\caption{Construction of the smooth,periodic piece-wise cubic spline with given inputs \(\{k_{\textrm{\scriptsize i}}, \alpha_{\textrm{\scriptsize i}} \}\)}
	\label{fig:spline_fit}
	\vspace{-1pt}
\end{figure}

\vspace{-10pt}
We construct the `bullet' profile (presented in the main text) of semi-major dimension \(r \) by using a spline-based shape parametrization, similar to~\citep{Rossinelli:2011b}. The piecewise-cubic spline (\cref{fig:spline_fit}) is fit in the polar coordinates (with restricted domain on \(\theta \in [0, \pi]\)), after choosing \(n\) control points and specifying their radial \( k_{\textrm{\scriptsize i}}\) and angular \( \alpha_{\textrm{\scriptsize i}}\) positions. We enforce periodicity and top--bottom symmetry by specifying zero-slope (clamped) boundary conditions for the half-spline and mirroring it about its central axis. Our freedom in the choice of \(n\) and consequently the set of \(\{k_{\textrm{\scriptsize i}}, \alpha_{\textrm{\scriptsize i}} \}\) (with cardinality/degrees of freedom = \(2n\)) renders it possible to get desirable shapes with high curvatures.

We tabulate the parameters used for constructing the `bullet' and other profiles used in the manuscript in \cref{tab:spline_params}. We note that across all splines, \( k_1\) and \(k_n\) (the scaling lengths of the diametrically opposite points) are identically set to \( 1.0\) to ensure invariance in the major dimensions of the shape. We then use this invariant dimension to define our \(\Rey\) and quantities derived from it.

\begin{table}[h!]
  \begin{center}
\def~{\hphantom{0}}
	\begin{tabular}{lccc}
		Spline ref.  & \(n\) & \(\{ k_{\textrm{\scriptsize i}}\}\) & \(\{ \alpha_{\textrm{\scriptsize i}}\}\) (in \(^\textrm{{\scriptsize o}}\))\\
		Bullet & 7 & \(\{1.00, 1.40, 1.42, 1.39, 0.72, 0.95, 1.00\}\) & \(\{43.0, 1.0, 1.0, 35.0, 90.0, 10.0\}\) \\
		figure 6e (in main text) & 6 & \(\{ 1.00, 1.28, 1.31, 1.29, 1.00, 1.00 \}\) & \( \{ 41.5, 1.75, 1.75, 60.0, 75.0 \} \)\\
		figure 6f (in main text) & 6 & \(\{1.00, 1.40, 1.42, 1.40, 0.99, 1.00\}\) & \( \{ 43.0, 1.0, 1.0, 45.0, 90.0 \} \)\\
  \end{tabular}
	\vspace{-10pt}
  \caption{Parameters for the splines used in the manuscript}
  \label{tab:spline_params}
  \end{center}
	\vspace{-25pt}
\end{table}

\vspace{-25pt}
\section{Three dimensional viscous flow mediated interactions--Numerics}
\vspace{-10pt}
Here we extend our approach of leveraging streaming for transport in three-dimensional simulations. We simulate both the master and slave as a spheres of diameter $ D_{\textrm{\scriptsize m}} =  0.025$m and $ D_{\textrm{\scriptsize s}} =  0.00625$m in an unbounded domain with uniform spacing of 1/2048 m in each dimension. The master is initialized at $ \boldsymbol{x}^0_{\textrm{\scriptsize m}} = [0.2, 0, 0] $ m and moves laterally with a constant velocity of $ U_{\textrm{\scriptsize l}} = |2D_{\textrm{\scriptsize m}}|$ms$ ^{-1} $. The slave is initialized at a surface-to-surface distance of $ 0.1D_{\textrm{\scriptsize m}} $ m behind the master at $ \boldsymbol{x}^0_{\textrm{\scriptsize s}} = [0.182, 0, 0] $ m. The simulation is set to stop at non-dimensional end time $ T = 2U_{\textrm{\scriptsize l}}t/D_{\textrm{\scriptsize m}} = 24 $. The viscosity $ \nu $ is set based on the linear motion Reynolds number $ \Rey = U_{\textrm{\scriptsize l}} D_{\textrm{\scriptsize m}}/\nu $. We oscillate the master with a fixed amplitude $ A = \thalf \epsilon D_{\textrm{\scriptsize m}}$, where $ \epsilon = 0.05$, with an angular frequency $ \omega $ determined by the oscillatory Reynolds number $ R_{\textrm{\scriptsize o}}  = \zeta \Rey = \epsilon \omega D_{\textrm{\scriptsize m}}/ (2\nu) $. We note that both master and slave are density matched, with $ \rho  = 1$ kgm$^{-3}$.

\begin{figure}[h!]
	\centerline{\includegraphics[width=1.0\linewidth]{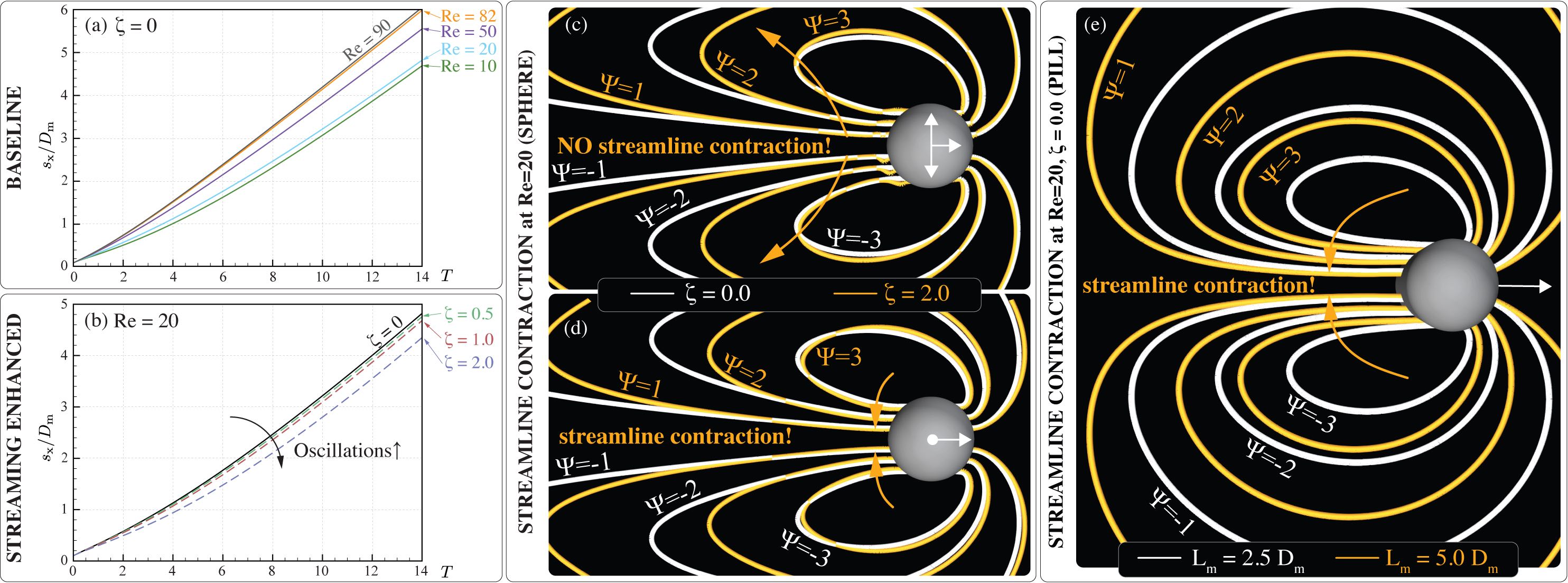}}
	\vspace{-10pt}
    \caption{ Separation distance \( s/D_{\textrm{\scriptsize m}} \) versus non-dimensional time \( T = 2 U_{\textrm{\scriptsize l}}t/D_{\textrm{\scriptsize m}}\) for (a) baseline transport at different \(\Rey\) and (b) streaming-enhanced transport with different \(\zeta\) at \(\Rey = 20\). Comparing the streamlines for the case using spherical master without (\(\zeta\) = 0, white streamlines) and with (\(\zeta\) = 2, orange streamlines) oscillation at Re = 20 reveals that streamline contraction does not happen on the (c) oscillation plane and barely occur on the (d) plane perpendicular to the oscillation plane ( \(\psi\) values scaled-up by 1250). Comparing the cases using linearly-translating pill-shaped master of different lengths \(L_{\textrm{\scriptsize m}}\) without oscillation reveals streamline contraction.}
	\vspace{-5pt}
    \label{fig:sphere}
\end{figure}

We conduct a baseline study on the performance of the spherical master across different $\Rey$ and observe better transport as $\Rey$ decreases (\cref{fig:sphere}a), but no trapping is achieved. We proceed to examine the usefulness of streaming effects by introducing oscillations ($\zeta = 0.5, 1.0, 2.0$) for the representative case of $\Rey = 20$. While \cref{fig:sphere}b suggests slight improvement in transport with increasing $\zeta$, it is evident that no trapping occurs. Looking into the streamlines of the cases without ($\zeta = 0$) and with ($\zeta=2.0$) oscillations, we see there is barely any streamline contraction in the oscillation plane (\cref{fig:sphere}c) and while slightly more streamline contraction occurs on the plane perpendicular to the plane of oscillation (\cref{fig:sphere}d), the acceleration in the wake flow provided is only capable of facilitating transport slightly and does not lead to trapping.

We elongate the master sphere to form a pill-shaped master of $D_{\textrm{\scriptsize m}}$ with hemispherical caps having an end-to-end length $ L_{\textrm{\scriptsize m}}$. At $\Rey = 20$, pill set in linear motion shows improvement in transport and eventually trapping as $L_{\textrm{\scriptsize m}}$ increases. We notice that mere increment in $L_{\textrm{\scriptsize m}}$ promotes streamline contraction in systems of linearly-translating pill (\cref{fig:sphere}e), and we can further enhance it by introducing oscillations.

\bibliographystyle{unsrtnat}

\end{document}